\documentclass[useAMS,usenatbib]{mn2e}
\usepackage{graphicx}
\usepackage{fixltx2e}
\usepackage{amsmath}
\usepackage{amssymb}

\newcommand{\ms}{$\,{\rm M}_\mathrm{\odot}$}

\newcommand{\diff}[2]{\frac{\partial (#1)}{\partial #2}} 
\newcommand{\diffb}[2]{\frac{\partial #1}{\partial #2}} 
\newcommand{\difft}[2]{\frac{{\rm d} #1}{{\rm d} #2}}
\newcommand{\bi}[1]{\textbf{\textit{#1}}}

\title[Stellar evolution with a radiative dynamo]{Stellar evolution of massive stars with a radiative alpha--omega dynamo}
\author[A. T. Potter, S. M. Chitre, C. A. Tout]{Adrian~T.~Potter$^1$\thanks{E-mail: apotter@ast.cam.ac.uk}, Shashikumar~M.~Chitre$^2$ and Christopher~A.~Tout$^1$ \\
$^1$Institute of Astronomy, The Observatories, Madingley Road, Cambridge CB3 0HA\\
$^2$Centre for Excellence in Basic Sciences, University of Mumbai}

\begin{document}

\date{Accepted 2012 May 29. Received 2012 May 28; in original form 2012 April 3}

\maketitle

\begin{abstract}

Models of rotationally--driven dynamos in stellar radiative zones have suggested that magnetohydrodynamic transport of angular momentum and chemical composition can dominate over the otherwise purely hydrodynamic processes. A proper consideration of the interaction between rotation and magnetic fields is therefore essential. Previous studies have focused on a magnetic model where the magnetic field strength is derived as a function of the stellar structure and angular momentum distribution. We have adapted our one--dimensional stellar rotation code, {\sc rose}, to model the poloidal and toroidal magnetic field strengths with a pair of time--dependent advection--diffusion equations coupled to the equations for the evolution of the angular momentum distribution and stellar structure. This produces a much more complete, though still reasonably simple, model for the magnetic field evolution. Our model reproduces well observed surface nitrogen enrichment of massive stars in the Large Magellanic Cloud. In particular it reproduces a population of slowly--rotating nitrogen--enriched stars that cannot be explained by rotational mixing alone alongside the traditional rotationlly--enriched stars. The model further predicts a strong mass--dependency for the dynamo--driven field. Above a threshold mass, the strength of the magnetic dynamo decreases abruptly and so we predict that more massive stars are much less likely to support a dynamo--driven field than less massive stars.

\end{abstract}

\section{Introduction}

The study of rotation in the radiative zones of stars is strongly coupled with the evolution of magnetic fields. Observation of stellar magnetic fields is difficult but a number of magnetic O and B stars have been discovered \citep{Donati01,Donati02,Neiner03,Donati06,Donati06b,Grunhut11}. Combined with this, a number of chemically peculiar A and B stars (known as Ap and Bp stars respectively) with surface field strengths up to $20$kG have been identified \citep[e.g.][]{Borra78,Bagnulo04,Hubrig05}. We direct the reader to \citet{Mathys09} for a review. These large--scale fields tend to have simple geometries and there is debate over whether they arise from fossil fields present during a star's formation \citep{Cowling45, Alecian08} or from a rotationally--driven dynamo operating in the radiative zone of the star \citep{Spruit99,Maeder04}. In this paper we focus on the latter but we give consideration to whether a fossil field can be sustained throughout the stellar lifetime.

In low--mass stars, where the outer region is convective, magnetic fields are expected to be formed in a strong shear layer at the base of the convection zone and then transported to the surface by convection and magnetic buoyancy \citep{Nordhaus10}. In radiative zones there is no strong bulk motion to redistribute magnetic energy. In most dynamo models, magnetic flux is redistributed by  magneto--rotational turbulence \citep{Spruit02}. This turbulence is also responsible for driving the generation of large--scale magnetic flux. This is the $\alpha$--effect \citep[e.g.][]{Brandenburg01} which applies to both poloidal and toroidal components, although in rotating systems shear is generally more effective at producing toroidal field from the poloidal component and so the $\alpha$--effect is needed for the poloidal field only. The toroidal field is instead maintained by the conversion of poloidal field into toroidal field by differential rotation. This is commonly referred to as an $\alpha$--$\Omega$ dynamo \citep{Schmalz91}.

Because observed fields are potentially strong enough to affect chemical mixing and angular momentum transport, their inclusion in stellar evolution models is essential. Rotation itself is a likely candidate to drive dynamo mechanisms within a star and theoretical models \citep[e.g.][]{Spruit99} have predicted magnetic fields that can produce turbulent instabilities which dominate the transport of angular momentum. Whilst the purely hydrodynamic evolution of the angular momentum distribution in main--sequence stars has been considered extensively in the framework of one--dimensional stellar evolution calculations \citep[e.g.][]{Meynet00,Heger00}, magnetic fields have received far less attention \citep{Maeder04, Brott11b}. The evolution of the angular momentum distribution and magnetic field strength have a significant effect on the final fate of a star and its ejecta. 

Apart from causing chemical mixing, sufficiently strong magnetic fields are expected to cause magnetic braking that results in the rapid spin down of rotating magnetic stars \citep{Mathys04}. It has been suggested that magnetic fields might explain the existence of slowly--rotating, chemically peculiar stars in surveys of rotating stars \citep{Hunter09}. We include a model for magnetic braking based on that of \citet{ud-Doula02} and show the effects it has on the models of magnetic stars.

Many studies of magnetic fields in massive main--sequence stars consider the Tayler--Spruit dynamo mechanism \citep{Spruit02}. This model asserts that pinch--type instabilities \citep{Tayler73, Spruit99} arise in toroidal fields that drive magnetic turbulence that enforces solid--body rotation. The growth of instabilities is controlled by magnetic diffusion which ultimately determines the equilibrium strength of the field. This idea was built upon by \citet{Maeder04} who found that the Tayler--Spruit dynamo did indeed result in far less differential rotation than in solely hydrodynamic models. It was also incorporated in the work of \citet{Brott11b} who compared stellar evolution calculations based on the Tayler--Spruit dynamo with the data from the VLT--FLAMES survey of massive stars \citep[e.g.][]{Evans05,Evans06}. They found reasonable agreement between the observed and simulated samples \citep{Brott11}. However, \citet{Potter12} found equally good agreement between the data from the VLT--FLAMES survey and purely hydrodynamic models based on models of \citet{Heger00} and \citet{Meynet00}.

In the models of of \citet{Spruit02} and \citet{Maeder04}, the magnetic field is purely a function of the stellar structure and rotation. Whilst it feeds back on the system via turbulent diffusivities, the magnetic field doesn't appear as an independent variable within the system. In this work we have continued along similar lines to \citet{Spruit02} but have developed a magnetic model where the poloidal and toroidal components are evolved via advection--diffusion equations derived from the induction equation. These are similar in form to the angular momentum evolution equation. The magnetic field and angular momentum evolution are coupled by turbulent diffusivities, magnetic stresses and conversion of poloidal field into toroidal field by differential rotation. The dynamo is completed by regeneration of magnetic flux by a simple $\alpha$--$\Omega$ dynamo. We look at how the predicted surface magnetic field varies with age and rotation rate for a range of initial masses and how a simulated population of magnetic stars compares to the data from the VLT--FLAMES survey of massive stars \citep{Dufton06}. We also consider how our model behaves with a strong initial fossil field but without the action of a dynamo.

In section~\ref{sec.model} we briefly review the model we use to simulate the magnetic fields including the equations for the $\alpha$--$\Omega$ dynamo and magnetic braking. In section~\ref{sec.results} we look at the predictions of the model for a range of stellar masses and initial rotation rates and how simulated populations compare with observations, in section~\ref{sec.discussion} we present a discussion of the results and in section~\ref{sec.conclusions} we give our concluding remarks.

\section{Rotating magnetic model}
\label{sec.model}

In order to simulate the magnetic field in stellar interiors we build on the code {\sc rose} described in \citet{Potter11} for one--dimensional stellar evolution calculations which include purely hydrodynamic angular momentum evolution.  The code is based on the Cambridge stellar evolution code {\sc stars} \citep{Eggleton71,Pols95,Eldridge09} and incorporates a number of different models for stellar rotation \citep[e.g.][]{Talon97,Heger00,Maeder03} as described in \citet{Potter11}. The evolution of the angular momentum is based on the shellular rotation hypothesis \citep{Zahn92} and treats the angular momentum evolution with a one--dimensional advection--diffusion equation. 

\subsection{Magnetic field evolution}

We approach the evolution of the magnetic fields in a similar way to the evolution of the angular momentum distribution. In the radiative zones of stars, turbulence from purely rotational or magnetorotational instabilities leads to the generation of magnetic field by an $\alpha\Omega$--dynamo mechanism. We assume a background velocity field of the form

\begin{equation}
\label{eq.velocity}
\textit{\textbf{U}}=\left\{U(r)P_2(\cos\theta),V(r)\frac{dP_2(\cos\theta)}{d\theta},\Omega(r)r\sin\theta\right\},
\end{equation}

\noindent where $P_2(x)$ is the 2nd Legendre polynomial and $U(r)$ and $V(r)$ are the components of the meridional circulation and are related by the continuity equation

\begin{equation}
V=\frac{1}{6\rho r}\frac{d}{dr}(\rho r^2U).
\end{equation}

\noindent The radial component, $U(r)$, is taken to be the same as described by \citet{Potter11} based on \citet{Maeder00}. It has been suggested that meridional circulation can be neglected in the presence of strong magnetic fields \citep{Maeder03b}. We discuss whether this is indicated by our model in section \ref{sec.angmom}. For now we leave it in our equations for completeness.

The evolution of the large--scale magnetic field is described by the induction equation

\begin{equation}
\label{eq.induction}
\diffb{\bi{B}}{t}=\nabla\times(\bi{U} \times \bi{B})-\nabla\times(\eta\nabla\times\bi{B}).
\end{equation}

\noindent Assuming an azimuthal form for the mean field we may write $\bi{B}$ as

\begin{equation}
\label{eq.magfield}
\bi{B}=B_{\phi}(r,\theta)\bi{e}_{\phi}+\nabla\times(A(r,\theta)\bi{e}_{\phi}).
\end{equation}

\noindent Substituting equations (\ref{eq.velocity}) and (\ref{eq.magfield}) into (\ref{eq.induction}) gives

\begin{align}
\label{eq.toroidalfield}
\diffb{B_{\phi}}{t}=&\,rB_r\sin\theta\diffb{\Omega}{r}+B_{\theta}\sin\theta\diffb{\Omega}{\theta}-\nonumber\\
& \frac{1}{r}\diffb{}{\theta}\left(V(r)\difft{P_2(\cos\theta)}{\theta}B_{\phi}\right) -\nonumber\\
&\frac{1}{r}\diffb{}{r}\left(rU(r)P_2(\cos\theta)B_{\phi}\right)-\nonumber \\
&\left(\nabla\times(\eta\nabla\times\bi{B})\right)_{\phi}
\end{align}
\noindent and
\begin{align}
\label{eq.poloidalfield}
\diffb{A}{t}=&\,-\frac{2V(r)}{r}\difft{P_2(\cos\theta)}{\theta}A\cot{\theta}-\nonumber\\ & \frac{U(r)P_2(\cos\theta)}{r}\diffb{Ar}{r}\sin\theta+\nonumber\\ & \alpha B_{\phi}-\nabla\times(\eta\nabla\times A\bi{e}_{\phi}),
\end{align}

\noindent where we have introduced the $\alpha$--term in equation (\ref{eq.poloidalfield}) to describe the regeneration of the poloidal field by the dynamo \citep{Schmalz91}. The radial and latitudinal components of the magnetic field are $B_r$ and $B_{\theta}$ respectively. Under the assumption of shellular rotation, the term $B_{\theta}\partial\Omega/\partial\theta\sin\theta=0$.

In order to reduce the equations to one dimension we need to choose the $\theta$--dependence of the magnetic field and perform a suitable latitudinal average of equations (\ref{eq.toroidalfield}) and (\ref{eq.poloidalfield}). First we choose $A(r,\theta)=\tilde{A}(r)\sin\theta$ so that in the limit of no meridional circulation or magnetic stresses, the poloidal field tends towards a dipolar geometry. Under this assumption $B_r=2\tilde{A}\cos\theta/r$ and $B_{\theta}=-{\rm d}(r\tilde{A})/{\rm d}r\sin\theta$. We could equally choose a quadrupolar or higher order geometry but we start with this as the simplest case. The radial field has negative parity about the equator so this must also be true of the toroidal field. The toroidal field must also vanish at the poles to avoid singularities. We therefore choose $B_{\phi}=\tilde{B_{\phi}}(r)\sin(2\theta)$. Again, this is not a unique choice but is the lowest order Fourier mode that meets our requirements. Finally we take $\alpha=\tilde{\alpha}(r)$ and $\eta=\tilde{\eta}(r)$. 

We take the average of a quantity $q$ to be

\begin{equation}
\langle q \rangle = \int^{\pi/2}_0q\sin\theta{\rm d}\theta = -\int^{\pi}_{\pi/2}q\sin\theta {\rm d}\theta.
\end{equation}

\noindent The second identity holds because of our choice of parity for the various terms in equations~(\ref{eq.toroidalfield}) and~(\ref{eq.poloidalfield}). Hereinafter we drop the use of angled brackets and write $q=\tilde{q}$ for the radially--dependent components of the magnetic field and related quantities. Taking averages of equations~(\ref{eq.toroidalfield}) and~(\ref{eq.poloidalfield}) we get

\begin{equation}
\label{eq.toroidalfield2}
\diffb{B_{\phi}}{t}=A\diffb{\Omega}{r}-\frac{6}{5r}VB_{\phi}-\frac{1}{10r}UB_{\phi}+r\diffb{}{r}\left(\frac{\eta}{r^4}\diffb{}{r}(r^3B_{\phi})\right)
\end{equation}
\noindent and
\begin{equation}
\label{eq.poloidalfield2}
\diffb{A}{t}=\frac{3V}{2r}A-\frac{U}{8r}\diffb{}{r}(Ar)+\frac{8\alpha}{3\pi} B_{\phi}+\diffb{}{r}\left(\frac{\eta}{r^2}\diffb{}{r}(r^2A)\right).
\end{equation}

\noindent In the case where diffusion dominates, $A\to 1/r^2$ and $B_{\phi}\to 1/r^3$. This is what we expect for a dipolar field. Our boundary conditions are $B_{\phi}=0$ and $B_{\theta}\propto\partial(rA)/\partial r=0$ at $r=0$ and $R_{*}$.

\subsection{Evolution of the angular momentum distribution}
\label{angmommod}

In the Taylor--Spruit dynamo \citet{Spruit02} angular momentum transport is driven by the Maxwell stress produced by the magnetic field. This process is assumed diffusive and an effective diffusion coefficient is derived. We treat the angular momentum evolution in radiative zones by extending equation~(12) of \citet{Potter11} to

\begin{align}
\label{eq.magamtransport}
\diff{r^2\Omega}{t}=&\,\frac{1}{5\rho r^2}\diff{\rho r^4\Omega U}{r}+\frac{3r}{8 \pi\rho}\langle\left(\nabla\times \bi{B}\right)\times\bi{B}\rangle_{\phi}+\nonumber\\ & \frac{1}{\rho r^2}\frac{\partial}{\partial r}\left(\rho D_{\rm tot}r^4\frac{\partial\Omega}{\partial r}\right),
\end{align}

\noindent where the pre--factor in the magnetic stress term comes from the combination of a factor of $1/4\pi$ for the permeability of free space and $3/2$ from the spherical average, $\langle r^2\sin^2\theta\rangle$, on the left--hand side. The term $D_{\rm tot}$ is the total diffusion of angular momentum that arises from a combination of purely rotationally--driven turbulence, magneto--rotational turbulence and convection. Purely hydrodynamic turbulence comes from Kelvin--Helmholtz instabilities that are driven by shear. We refer to this diffusion coefficient as $D_{\rm KH}$. There are other sources of hydrodynamic turbulence, including an effective diffusion owing to the meridional circulation, but we shall group these all in $D_{\rm KH}$.  We use the formulation of \citet{Potter11}, based on that of \citet{Maeder03}, but other formulations may be used instead, described by \citet{Potter12}. The diffusion by convective transport is $D_{\rm con}$ and is based on the effective diffusion from mixing--length theory \citep{Bohm58}. Finally the magnetic diffusion is $D_{\rm mag}$. With this notation $D_{\rm tot}=D_{\rm KH}+D_{\rm con}+D_{\rm mag}$. After averaging the magnetic stress term in equation (\ref{eq.magamtransport}) over co--latitude we find

\begin{align}
\diff{r^2\Omega}{t}=&\,\frac{1}{5\rho r^2}\diff{\rho r^4\Omega U}{r}+\frac{3}{64 \rho r^3 B_\phi}\diffb{}{r}\left(r^3B^2_{\phi}A\right)+\nonumber\\ & \frac{1}{\rho r^2}\frac{\partial}{\partial r}\left(\rho D_{\rm tot} r^4\frac{\partial\Omega}{\partial r}\right),
\end{align}

\noindent where a factor of $8/\pi$ appears in the Maxwell stress term owing to the spherical average.

 We see that the Maxwell stress does not act diffusively as is often suggested. \citet{Spruit02} equates the Maxwell stress, $S$, to $r \rho\nu_{\rm e}\partial\Omega/\partial r$, where $\nu_{\rm e}$ is some effective diffusivity. This automatically assumes that the large scale stresses lead to solid body rotation and is unjustified. It leads to a diffusion coefficient of the form $\nu_{\rm e}\propto(\partial\Omega/\partial r)^{-1}$ and so high diffusion rates for small shear. We could have equally assumed any similar relation such as $S=(\rho\hat{\nu}_{\rm e}/r)\partial(r^2\Omega)/\partial r$, where $\hat{\nu}_{\rm e}$ is now an effective diffusivity which drives the system towards uniform specific angular momentum. For \citet{Spruit02} this never becomes a problem because he assumes a steady--state saturated magnetic field but it does present a problem for systems where the magnetic field strength is independently derived. The magnetic stress term in fact acts advectively and so can increase the amount of shear in the system.

\subsection{Magnetic diffusion}

Instead of relying on the large scale Maxwell stress to redistribute angular momentum in radiative zones, we use the magnetic turbulence from the Tayler--instability \citep[][]{Tayler73}. Turbulent diffusion coefficients for this instability were proposed by \citet{Spruit02} and \citet{Maeder04}. We follow a similar method to derive the associated diffusion coefficients here. The main difference is that we solve for the magnetic field and hence the Alfv\'{e}n velocity independently instead of treating it as a function of the rotation rate.

First, the energy of the instability must be enough to overcome the restoring buoyancy force. This puts a limit on the vertical extent of the magnetic instability

\begin{equation}
\label{eq.lr}
l_r<\frac{r\omega_{\rm A}}{N},
\end{equation}

\noindent where $\omega_{\rm A}^2\approx B_{\phi}^2/4\pi r^2 \rho$ is the Alfv\'{e}n frequency and $N$ is the relevant buoyancy frequency. If this length scale is too small then the magnetic diffusivity damps the instability. \citet{Spruit02} takes this limit to be

\begin{equation}
\label{eq.lr2}
l_r^2>\frac{\eta\Omega}{\omega_{\rm A}^2}.
\end{equation}

\noindent When account is taken of the thermal diffusivity, the buoyancy frequency given by \citet{Maeder04} is

\begin{equation}
\label{eq.bvfrequency}
N^2=\frac{\eta/K}{\eta/K+2} N_T^2 + N_{\mu}^2,
\end{equation}

\noindent where $K$ is the thermal diffusivity, $N_T^2$ is the Brunt--V\"{a}is\"{a}l\"{a} frequency and $N_{\mu}^2$ is the frequency associated with the mean molecular weight gradient. Substituting equation (\ref{eq.bvfrequency}) into equations (\ref{eq.lr}) and (\ref{eq.lr2}) gives a quadratic equation for $\eta$,

\begin{equation}
\label{eq.quadratic}
(N_T^2+N_{\mu}^2)\eta^2+\left(2KN_{\mu}^2-\frac{r^2\omega_{\rm A}^4}{\Omega}\right)\eta-2Kr^2\omega_{\rm A}^4=0.
\end{equation}

In the limit $N_{\mu}^2\gg N_T^2$ and $K\ll\eta$ we recover equation~(1) of \citet{Maeder04} and in the limit $N_{\mu}^2\ll N_T^2$ and $K\gg\eta$ we recover their equation~(2). In most cases we find that $K\gg\eta$ and $N_T^2\gg N_{\mu}^2$ in which case we get

\begin{equation}
\label{eq.etaapprox}
\eta\approx r^2\Omega\left(\frac{\omega_{\rm A}}{\Omega}\right)^2\left(\frac{\Omega}{N}\right)^{1/2}\left(\frac{K}{r^2 N_T}\right)^{1/2}.
\end{equation}

\noindent In equation (\ref{eq.quadratic}) we make the substitution $\eta=C_{\rm m}\eta'$ where $C_{\rm m}$ is a calibration constant which we expect to be of order unity. The chemical composition of the star evolves in radiative zones according to the equation

\begin{equation}
\label{eq.magchemevolution}
\frac{\partial X_i}{\partial t}=\frac{1}{r^2}\frac{\partial}{\partial r}\left({\rm Pr_c}D_{\rm tot}r^2\frac{\partial X_i}{\partial r}\right),
\end{equation}

\noindent where ${\rm Pr_c}$ is the turbulent chemical Prandtl number and $X_i$ is the mass fraction of element $i$. Similarly we take the magnetic diffusivity to be $\eta={\rm Pr_m}\,D_{\rm mag}$ where ${\rm Pr_m}$ is the turbulent magnetic Prandtl number. We look at the effect of varying these two parameters in section \ref{sec.calibration} but we expect the magnetic Prandtl number to be of order unity \citep{Yousef03}.

\subsection{Dynamo model}

We describe the dynamo generation parameter by taking $\alpha=\gamma r/\tau_{\rm a}$ where $\gamma$ is an efficiency parameter and $\tau_{\rm a}$ is the amplification time scale of the field. Following \citet{Maeder04} we take $\tau_{\rm a}=N/\omega_{\rm A}\Omega q$ where $q=\partial(\log\Omega)/\partial(\log r)$. Combining these our dynamo efficiency is given by

\begin{equation}
\label{eq.alpha}
\alpha=\gamma\frac{r\omega_{\rm A}\Omega q}{N}.
\end{equation}

\subsection{Magnetic braking}

Strongly magnetic intermediate--mass stars typically have rotation rates much slower than other stars in their parent population \citep{Mathys04}. If the Alfv\'{e}n radius, the radius at which the magnetic energy density is the same as the kinetic energy density in the stellar wind, is larger than the stellar radius then magnetic braking allows additional angular momentum to be carried away by the stellar wind. Consider equation (\ref{eq.magamtransport}). Writing $\int_0^m{\rm d}m=\int_0^R4\pi r^2 \rho {\rm d}r$ we obtain the boundary condition for angular momentum loss from the surface

\begin{equation}
\difft{H_{\rm tot}}{t}=4\pi R^4\rho D_{\rm tot}\left(\diffb{\Omega}{r}\right)_{R}
\end{equation}

\noindent where ${\rm d}H_{\rm tot}/{\rm d}t$ is the total rate of angular momentum loss from the star and is given by

\begin{equation}
\difft{H_{\rm tot}}{t}=R_A^2\Omega\dot{M}=\sigma^2J_{\rm surf}.
\end{equation}

\noindent The Alfv\'{e}n radius is $R_A$, $\sigma=R_A/R$ and $J_{\rm surf}$ is the specific angular momentum at the surface of the star. Following the analysis of \citet{ud-Doula02} we can calculate the magnetic efficiency

\begin{equation}
\label{eq.eta}
\phi(r)=\frac{B_*^2R^2}{\dot{M}v_{\infty}}\frac{(\frac{r}{R})^{-4}}{1-\frac{R}{r}},
\end{equation}

\noindent where $v_{\infty}=v_{\rm esc}=\sqrt{2g_{\rm eff}R}$ and $v_{\rm esc}$ is the escape velocity at the stellar surface. We have assumed that the external field is dipolar ($q=3$). The Alfv\'{e}n radius is typically taken where the dynamo efficiency equals unity. Rearranging equation (\ref{eq.eta}), and setting $\phi=1$ and $\sigma=r/R=R_A/R$ at $r=R_A$ we find

\begin{equation}
\label{eq.sigma4}
\sigma^4-\sigma^3=\frac{B_*^2R^2}{\dot{M}v_{\rm esc}}.
\end{equation}

\noindent We assume $\sigma\gg 1$ so that

\begin{equation}
\label{eq.sigma2}
\sigma^2=\sqrt{\frac{B_*^2R^2}{\dot{M}v_{\rm esc}}}
\end{equation}

\noindent for the remainder of this paper. If $R_A<R$ then we take $\sigma=1$ so that, as star loses mass, material carries away the specific angular momentum of the material at the surface. When we approach this limit we should calculate $\sigma$ exactly from (\ref{eq.sigma4}) but for now we assume that (\ref{eq.sigma2}) remains valid. In section~\ref{sec.mvb} we typically find either strong fields where $\sigma\gg 1$ or very weak fields where we can safely take $\sigma=1$. So far we have been unable to produce a stable model for the mass--loss rates of \citet{Vink01} and so use the rate of \citet{Reimers75} in equation~(\ref{eq.sigma2}). For intermediate--mass stars on the main sequence this approximation is reasonably accurate.

\subsection{Free parameters}

Like most theories for stellar rotation and magnetic field evolution we have produced a closed model which depends on a number of free parameters. We look at typical physically motivated values for these parameters in section~\ref{sec.calibration} and also the effect of varying them. In total we have four free parameters. The parameter $C_{\rm m}$ affects the overall strength of the turbulent diffusivity. The magnetic and chemical Prandtl numbers, ${\rm Pr_m}$ and ${\rm Pr_c}$, describe how efficiently the turbulent diffusivity transports magnetic flux and chemical composition compared to angular momentum. And $\gamma$ affects the strength of the dynamo generation. Whilst ${\rm Pr_m}$ and $C_{\rm m}$ are both expected to be of order unity we have left them as free parameters for the moment to maintain of generality.

\section{Results}
\label{sec.results}

\begin{figure}
\begin{center}
\includegraphics[width=0.49\textwidth]{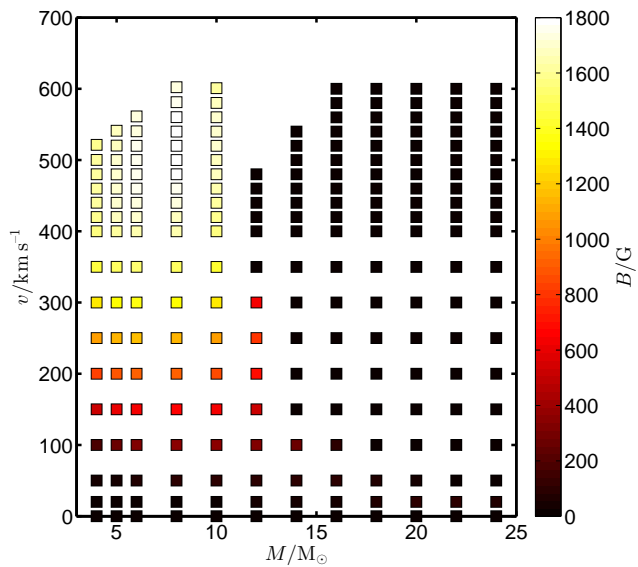}
\end{center}
\caption[Grid of models used in section \ref{sec.results}]{Grid of models considered in section \ref{sec.results}. The colour of each point indicates the surface field strength at the zero--age main sequence.}
\label{fig.grid}
\end{figure}

We simulated a grid of models with masses $4<M/$\ms$<24$ and initial rotation rates $0<v_{\rm ini}/{\rm km\,s^{-1}}<600$, except where the initial rotation rate is greater than the critical rotation rate of the star. All of the models described are at LMC metallicity as used by \citet{Brott11b}. We set $C_{\rm m}=1$ and ${\rm Pr_m}=1$. We also set $\gamma = 10^{-15}$ which results in a maximum field strength across the whole population of $B\approx 20\,{\rm kG}$. The maximum terminal--age main--sequence (TAMS) nitrogen enrichment in the simulated magnetic population, including observational constraints, is matched with the maximum enrichment in the slowly rotating population of \citet{Hunter09}. This gives ${\rm Pr_c}=0.01$. In each model the rotation and magnetic field were allowed to relax to equilibrium at the zero--age main sequence (ZAMS). The grid of initial models is shown in Fig.~\ref{fig.grid} which also shows the ZAMS surface field strength in each model. We will look at this in more detail in section~\ref{sec.mvb}.

\subsection{Magnetic field evolution}
\label{sec.evolution}

\begin{figure*}
\begin{center}
\hspace*{-0.5cm}
\includegraphics[width=0.49\textwidth]{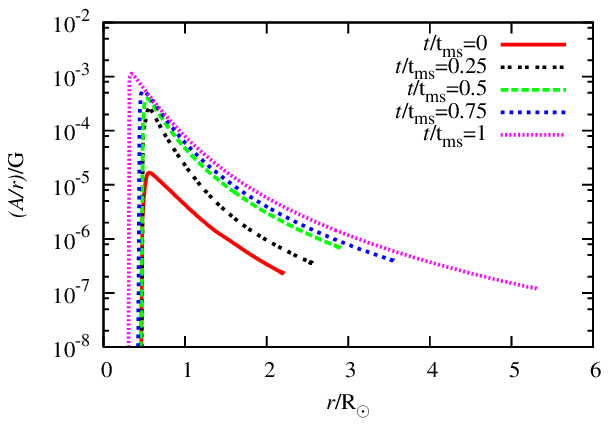}\includegraphics[width=0.49\textwidth]{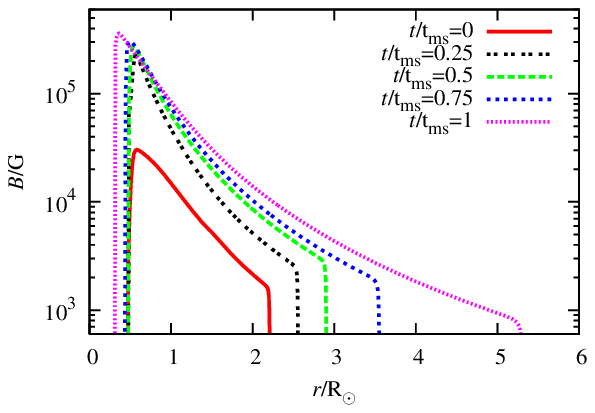}
\end{center}
\caption[Magnetic field evolution in a $5$\ms\ star initially rotating at $300\,{\rm km\,s^{-1}}$ without magnetic braking]{Evolution of the magnetic field in a $5$\ms\ star initially rotating at $300\,{\rm km\,s^{-1}}$ without magnetic braking. The left plot shows the magnetic potential for the poloidal field and the right plot shows the toroidal field. The $\alpha$--effect produces a weak poloidal field which is efficiently converted into toroidal field by differential rotation. For each component, the field strength is approximately three orders of magnitude smaller at the surface than the core. The ratio of the toroidal and poloidal field strengths is of the order~$10^9$.}
\label{fig.mag1}
\end{figure*}

Owing to the strong magnetically--induced turbulence, the toroidal field behaves roughly as $B_{\phi}\propto r^{-3}$ and the poloidal field behaves as $A\propto r^{-2}$ so both are much stronger towards the core than at the surface of the star as shown in Fig. \ref{fig.mag1}. The toroidal field falls to zero within a very narrow region near the surface of the star to meet the boundary conditions. The strength of the toroidal field predicted is around nine orders of magnitude larger than the poloidal field. This is because the $\Omega$--effect, the conversion of poloidal field into toroidal field by differential rotation, is much stronger than the $\alpha$--effect which regenerates the poloidal field. We take the surface value of the field to be the strength of the toroidal field just below the boundary layer. If we were instead to take the poloidal field, we would need a larger value of $\gamma$ to produce a stronger field. In this case the toroidal field is around six orders of magnitude larger than the poloidal field. So a surface poloidal field of $10^3\,{\rm G}$ would correspond to a toroidal field of $10^9\,{\rm G}$ just below the surface. The fields then increase by several orders of magnitude towards the core. Not only do these field strengths seem unreasonably energetic but also the magnetic stresses result in cores that are spinning near or above break--up velocity. However, spectropolarimetric observations have concluded that the large--scale structure of the external magnetic fields of massive stars are largely dipolar so there must be some mechanism for converting the toroidal field into poloidal field at the surface. It is likely that the stellar wind stretches the field lines in the radial direction, changing the toroidal field to a radial geometry as material is ejected from the stellar surface \citep{Parker58}.

Owing to the very large value for $D_{\rm con}$ predicted from mixing--length theory, the predicted field is extremely weak within the convective core. This is somewhat at odds with our observations in the Sun where large--scale magnetic flux can be transported through a convective region without being destroyed. It may be that convection is better treated by an anisotropic diffusivity. Certainly in the Sun, where the outer envelope is convective, we see latitudinal variations in the surface angular velocity which we have ignored owing to the shellular rotation hypothesis \citep{Zahn92} which applies to the radiative zones of massive stars. Therefore this does not strongly affect our model but deserves further consideration in the future.

\begin{figure}
\begin{center}
\hspace{-0.5cm}\includegraphics[width=0.49\textwidth]{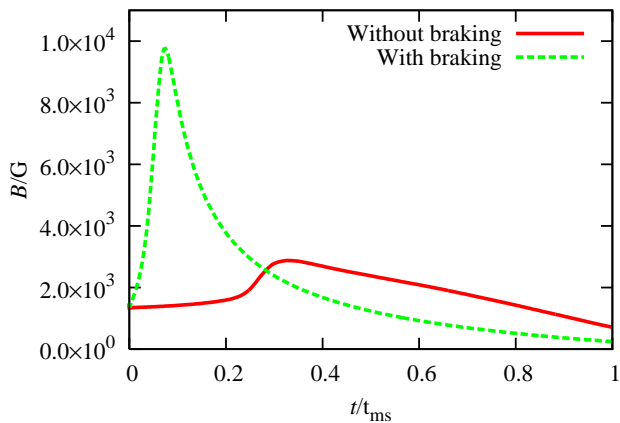}
\end{center}
\caption[Evolution of the magnetic field in a $5$\ms\ star initially rotating at $300\,{\rm km\,s^{-1}}$ with magnetic braking]{Evolution of the surface magnetic field strength in a $5$\ms\ star initially rotating at $300\,{\rm km\,s^{-1}}$ with and without magnetic braking. The surface field strength shows only a slight degree of variation during the main sequence when there is no magnetic braking. When magnetic braking is included the field strength peaks sharply after the ZAMS and then decays away rapidly. However, the field strength at the end of the main sequence is still several hundred Gauss.}
\label{fig.mag2}
\end{figure}

We first consider models in the absence of magnetic braking in order to distinguish evolutionary effects owing to the dynamo from those caused by braking. In this case, although the surface field only exhibits a small degree of variation (Fig. \ref{fig.mag2}), the magnetic field inside the star becomes significantly stronger during the course of the main sequence. The surface magnetic field reaches a peak strength and then weakens towards the end of the main sequence. However, this change is always within a factor of three of the ZAMS value.  This is consistent with the model of \citet{Tout96} in which Ae/Be stars tap rotational energy early in their lives. The enhancement of the field inside the star is largely because the Brunt--V\"{a}is\"{a}l\"{a} frequency decreases as the star expands during the main sequence. It is also partly because the amount of differential rotation increases as a result of the changing hydrostatic structure of the star.

We might intuitively expect that the spin down of the star owing to magnetic braking would cause the magnetic field to decay rapidly and this is true later in the life of the star. However, the inclusion of magnetic braking first leads to a significant enhancement of the magnetic field shortly after the ZAMS. When braking is included, the loss of angular momentum from the surface is so fast that diffusion of angular momentum cannot prevent a build up of shear within the radiative envelope. This drives additional generation of magnetic flux through the $\alpha$--$\Omega$ dynamo and actually causes a much stronger peak field than without magnetic braking. The magnetic diffusion eventually reduces the amount of differential rotation and the magnetic spin down results in a weaker dynamo and faster rate of field decay. However, the field remains sufficiently large throughout the main sequence that the rate of chemical transport is still large enough to cause a significant amount of nitrogen enrichment. We discuss this further in section~\ref{sec.hunter}. Although the eventual decay of the surface field in the presence of magnetic braking is quite rapid, the field strength at the end of the main sequence is still several hundred Gauss. This is consistent with the observation that all chemically peculiar Ap stars have strong fields \citep{Auriere07}.

\subsection{Effect on angular momentum distribution}
\label{sec.angmom}

\begin{figure}
\begin{center}
\hspace{-0.5cm}
\includegraphics[width=0.49\textwidth]{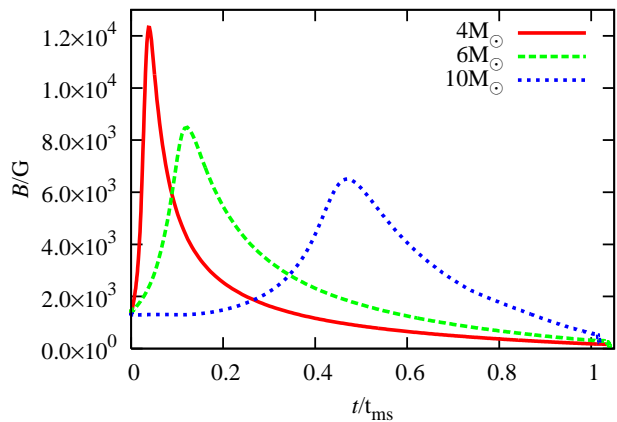}
\end{center}
\caption[Evolution of the surface magnetic field strengths in rotating stars of various masses]{Evolution of the surface magnetic field strengths in $4$\ms, $6$\ms\ and $10$\ms\ stars initially rotating at $300\,{\rm km\,s^{-1}}$ with an $\alpha$--$\Omega$ dynamo and magnetic braking. The maximum field strength is much greater in less massive stars. For all masses of star, the field strength increases sharply at the start of the main sequence owing to the rapid loss of angular momentum at the surface because of magnetic braking. This causes differential rotation which drives additional flux generation by the dynamo. This peak occurs later for more massive stars both in absolute time and as a fraction of their main--sequence lifetime. Following this, the field decays rapidly over the remainder of the main sequence.}
\label{fig.magstrength}
\end{figure}

\begin{figure*}
\begin{center}
\includegraphics[width=0.49\textwidth]{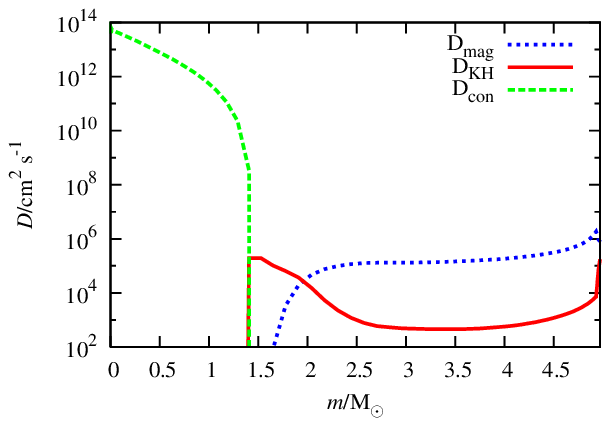}\includegraphics[width=0.49\textwidth]{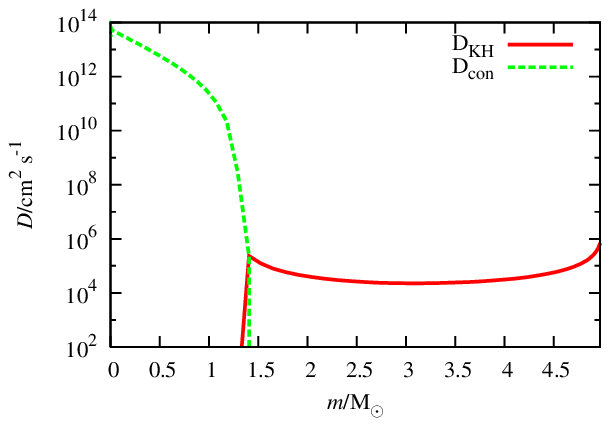}
\end{center}
\caption[Turbulent diffusivities governing angular momentum transport in a magnetic, rotating $5$\ms\ star]{Diffusivities for the angular momentum resulting from convection, hydrodynamic and magnetohydrodynamic effects in a $5$\ms\ star initially rotating at $300\,{\rm \, km\,s^{-1}}$. The left plot is for a magnetic star whereas the right plot is for a non--magnetic star. We note that the model predicts more efficient transport by magnetic effects compared to purely hydrodynamic effects. We also note that in the magnetic star, the diffusion of angular momentum by hydrodynamic turbulence is greatly reduced because the magnetic field reduces shear. There is a small region near the convective core where the magnetic diffusion becomes much smaller owing to mean molecular weight gradients. In this region the hydrodynamic turbulence dominates. This region only exists at the start of the ZAMS because the field becomes much stronger shortly after and the effects of rotation decrease as magnetic braking spins the star down.}
\label{fig.diffusion}
\end{figure*}

\begin{figure}
\begin{center}
\hspace{-0.5cm}
\includegraphics[width=0.49\textwidth]{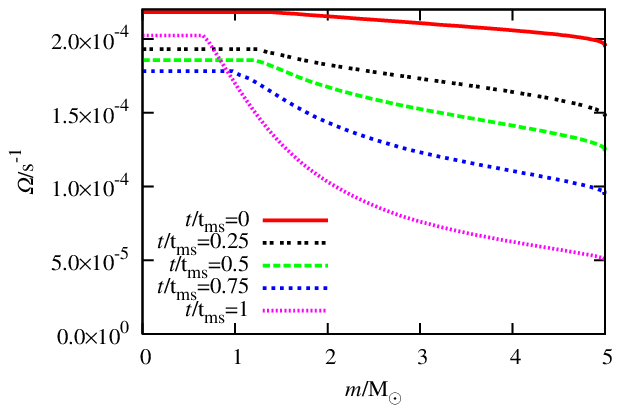}\\
\hspace{-0.5cm}
\includegraphics[width=0.49\textwidth]{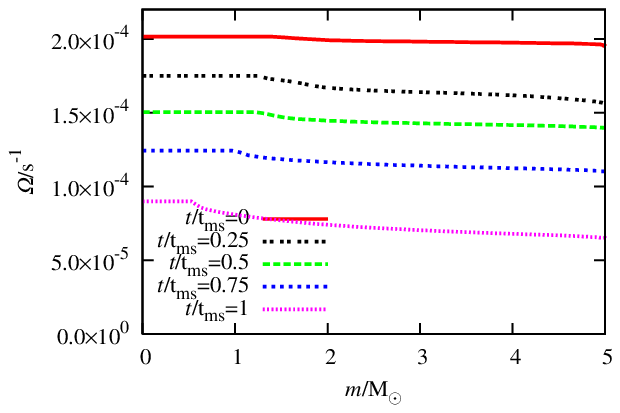}\\
\hspace{-0.5cm}
\includegraphics[width=0.49\textwidth]{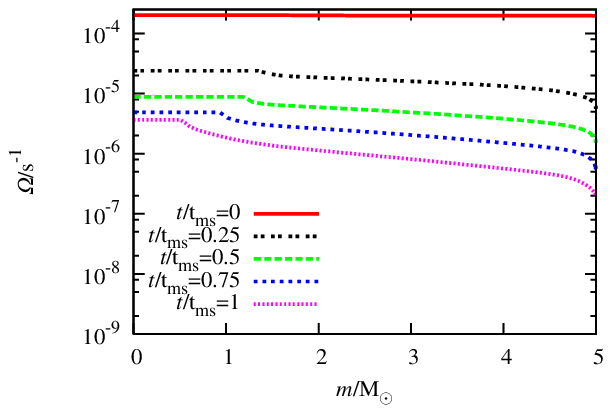} 
\end{center}
\caption[Evolution of the angular momentum distribution in a rotating, magnetic $5$\ms\ star]{Evolution of the angular momentum distribution in a $5$\ms\ star initially rotating at $300\,{\rm \, km\,s^{-1}}$. The top plot is for a non--magnetic star, the middle plot is for a magnetic star without braking and the bottom plot is for a magnetic star with braking. In magnetic stars without braking, the strong magnetic turbulence results in much less shear than the non--magnetic equivalent. Stronger diffusion in the magnetic stars also leads to far less differential rotation between the core and the envelope. This causes higher surface rotation in the non--braked magnetic star compared to the non--magnetic star. When braking is introduced to the magnetic star it spins down rapidly. The angular momentum loss from the surface leads to a much higher degree of differential rotation in the magnetic star with braking compared to the magnetic star without braking.}
\label{fig.omega}
\end{figure}

The effect of the magnetic field on the angular momentum distribution of a star has profound implications for its chemical evolution and the properties of its remnant. Shear arises in stars mostly as a result of changes in the structure from ongoing evolution and mass loss. Rotation also causes meridional circulation in stars. This contributes to the shear as we discussed in section~\ref{sec.angmom}. In the magnetic case, where magnetic braking is included, meridional circulation dominates over the magnetic stresses at the ZAMS for almost the entire star. For a $5$\ms\ star initially rotating at $300\,{\rm km\,s^{-1}}$ the meridional circulation is approximately six orders of magnitude stronger in the outer layers than the magnetic stresses at the ZAMS. Through most of the envelope the difference is between one and three orders of magnitude. However, when the magnetic field grows rapidly shortly after the ZAMS and magnetic braking begins to rapidly spin down the star this reverses and the magnetic stresses become much more important than the meridional circulation for the remainder of the main sequence. As we see in Fig~\ref{fig.magstrength}, this peak occurs later for more massive stars as a fraction of main--sequence lifetime and so the meridional circulation can dominate for longer. Therefore, whilst it is true that the meridional circulation has little effect on the evolution of magnetic stars for most of the main sequence, it is not necessarily true close to the ZAMS. This also suggests that the circulation terms in the induction equation, which we ignore in this paper, may in fact make a significant contribution to the transport of magnetic flux.

Apart from the physical effects that produce shear within the radiative envelope the other major factor that affects the angular momentum distribution is the strength of the turbulent diffusion. We have plotted the major diffusion coefficients at the ZAMS for a $5$\ms\ star initially rotating at $300\,{\rm km\,s^{-1}}$ in Fig.~\ref{fig.diffusion}. The overall diffusion coefficient predicted by the magnetic model is significantly larger than produced by hydrodynamic turbulence alone. We note that $D_{\rm KH}\propto\left(\partial\Omega/\partial r\right)^2$ predicted in the magnetic model is significantly lower than in the non--magnetic model. Whilst magnetic stresses should produce more shear than in the non--magnetic model, the diffusion coefficient is sufficiently high to cause an overall reduction in shear. This is illustrated in  Fig.~\ref{fig.omega} where we have plotted the evolution of the angular momentum distribution for the same star without a magnetic field, with a magnetic field but without braking and with both a magnetic field and braking. There is a small region near the convective core in the magnetic star where the magnetic diffusion becomes much smaller owing to mean molecular weight gradients. In this region the hydrodynamic turbulence dominates. This region only exists at the start of the ZAMS because the field becomes much stronger shortly after and the effects of rotation decrease as magnetic braking spins the star down.

In Fig.~\ref{fig.omega} we see that, in the magnetic star without braking, there is far less differential rotation throughout the star than in the non--magnetic star. This also means that the cores of magnetic stars are likely to be rotating more slowly than non--magnetic stars even before the effects of braking are included. When braking is included we see much the same trend except, in the model with magnetic braking, the whole star spins down rapidly. The typical Alfv\'{e}n radius for this star is approximately $50$, meaning that the rate of angular momentum loss is several thousand times faster than without a magnetic field. We note that there is far more differential rotation in this star compared with the magnetic star without braking. This is because of the rapid loss of angular momentum from the surface of the star.

\subsection{Mass--rotation relation of the main--sequence field strength}
\label{sec.mvb}

\begin{figure}
\begin{center}
\includegraphics[width=0.49\textwidth]{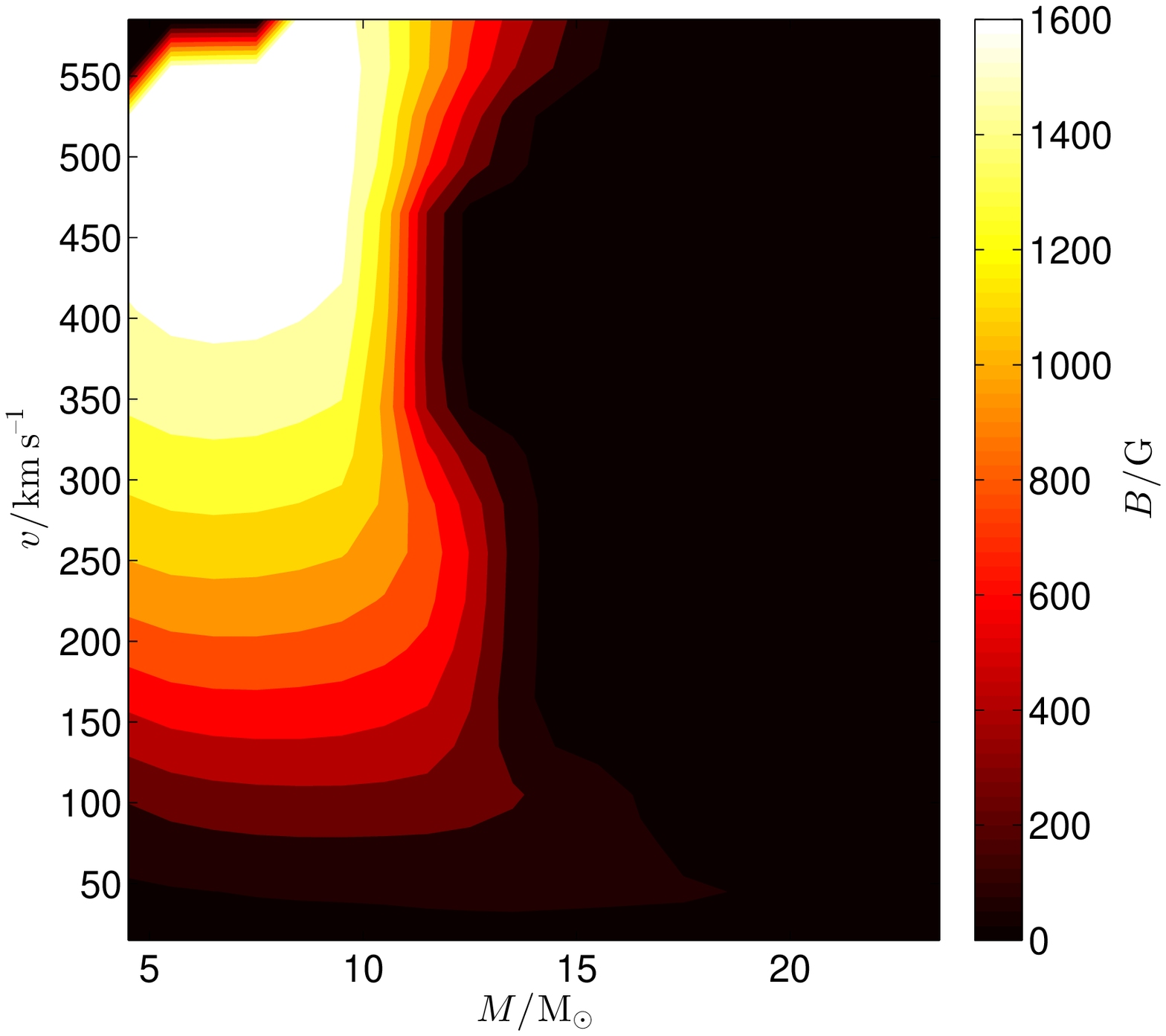}
\end{center}
\caption[Main--sequence magnetic field strengths for ZAMS stars with varying masses and rotation rates]{Main--sequence magnetic field strengths for intermediate--mass ZAMS stars at different rotation rates. Stars more massive than $15$\ms\ have almost no magnetic activity except for a weak field in slow rotators. The strongest fields occur in the most rapidly rotating stars with $4<M/$\ms$<10$.}
\label{fig.magevol}
\end{figure}

\begin{figure}
\begin{center}
\includegraphics[width=0.49\textwidth]{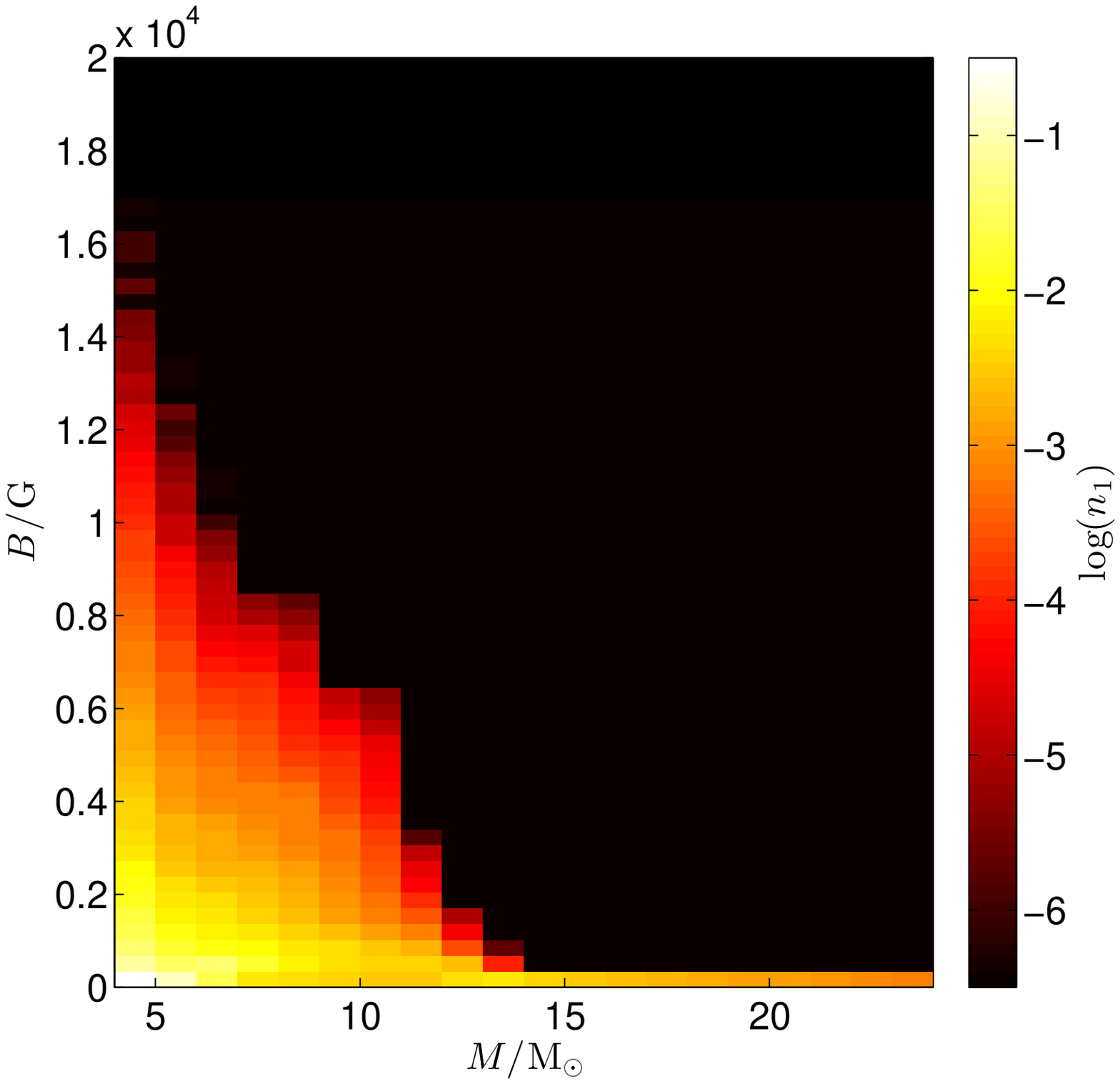}
\end{center}
\caption[Masses and magnetic field strengths of stars in a synthetic stellar population with ongoing star formation]{Simulated masses and magnetic field strengths for a population of stars drawn from the grid of models shown in Fig. \ref{fig.grid}. The population undergoes continuous star formation, is drawn from a Salpeter IMF and the velocity distribution is Gaussian with mean $\mu=145\,{\rm \, km\,s^{-1}}$ and standard deviation $\sigma=94\,{\rm \, km\,s^{-1}}$. The number of stars in each bin as a fraction of the total number of stars is $n_1$. We see that lower--mass stars support much stronger fields. There is very little magnetic activity in stars more massive than around $15$\ms.}
\label{fig.mb}
\end{figure}

Historically the presence of strong magnetic fields in massive stars has been thought to be mainly confined to A~stars and perhaps some of the lower--mass B stars \citep{Mathys09}. This may have been because of the difficulty in observing magnetic effects in the broad absorption features of more massive stars \citep{Petit11}. However, as the amount of available data has grown, thanks to surveys such as MiMeS project \citep[e.g.][]{Wade09}, it has become clear that this is not caused simply by selection effects.

By applying our model to the grid shown in Fig.~\ref{fig.grid} we are able to track the evolution of the surface field strength of stars and, in particular, how it varies with mass and rotation rate. We show this dependence at the ZAMS in Fig.~\ref{fig.magevol}. It is immediately apparent that although stars less massive than around $15$\ms\ are able to sustain significant fields, no significant field is predicted in more massive stars except in very slow rotators. Even for high--mass, slow rotators the field doesn't exceed $200{\rm \, G}$. The transition between a strong ZAMS field and no field is sharpest in rapid rotators. This transition is caused by the interaction between hydrodynamic and magnetic turbulence. If $D_{\rm KH}$ exceeds $D_{\rm mag}$ for a sufficiently large region of the radiative envelope, the magnetic field decays exponentially and cannot be sustained by the dynamo. Because $\alpha\propto\omega_A$, the strength of the dynamo weakens with the magnetic field. In the case where $D_{\rm mag}$ is the dominant turbulent process, this is matched by a greater reduction in the turbulent diffusivity because, for most of the envelope, $D_{\rm mag}\propto\omega_A^2$. As the diffusivity drops, the field is less efficiently dissipated and so an equilibrium is reached. When $D_{\rm KH}$ dominates and the field decays the diffusivity is largely unaffected and so the dynamo continues to weaken causing the field to completely disappear. At higher masses and rotation rates $D_{\rm KH}$ is larger and so catastrophic quenching occurs for lower dynamo efficiencies. Assuming that both instabilities act in the radiative envelope, this explains why magnetic fields are more likely to be observed in A~stars and less frequently in O and B stars. 

Given that the magnetic field strength increases sharply after the main sequence before decaying away exponentially as discussed in section~\ref{sec.evolution}, we consider the distribution of magnetic field strengths in a population of stars with a continuous distribution of ages. The population is shown in Fig.~\ref{fig.mb}. The population undergoes continuous star formation, is drawn from a Salpeter IMF and a Gaussian velocity distribution with mean $\mu=145\,{\rm km\,s^{-1}}$ and standard deviation $\sigma=94\,{\rm km\,s^{-1}}$. We see that magnetic activity is highest in the least massive stars. As before, stars more massive than around $15$\ms\ show no magnetic activity. We note that the stars with the strongest fields fall outside the observational limits of the VLT--FLAMES survey of massive stars \citep{Dufton06}. We discuss this further in section~\ref{sec.hunter}.

 We therefore predict two distinct populations of stars. The first is a population of slowly rotating, magnetic and chemically peculiar stars with masses less than $15$\ms. The second is a population of more massive stars that are non--magnetic and follow the trend discussed by \citet{Hunter09} and \citet{Potter12}, where rotation and nitrogen enrichment have a strong positive correlation. This is precisely what we observe \citep{Hunter09}. We may still observe A~stars that are rapidly rotating but not highly enriched. These stars should still support a strong magnetic field but are sufficiently young that no chemical enrichment has occurred. These rapidly rotating stars would be very infrequent owing to the efficient spin down by the magnetic braking. A rapidly rotating, highly magnetic massive star was observed by \citet{Grunhut12}. The star has a mass of $5.5$\ms\ and a surface rotation velocity of $290\,{\rm km\,s^{-1}}$ but has a surface field strength in excess of $10\,$kG.

\subsection{Effect on the Hertzsprung--Russel diagram}

\begin{figure}
\begin{center}
\hspace{-0.5cm}
\includegraphics[width=0.49\textwidth]{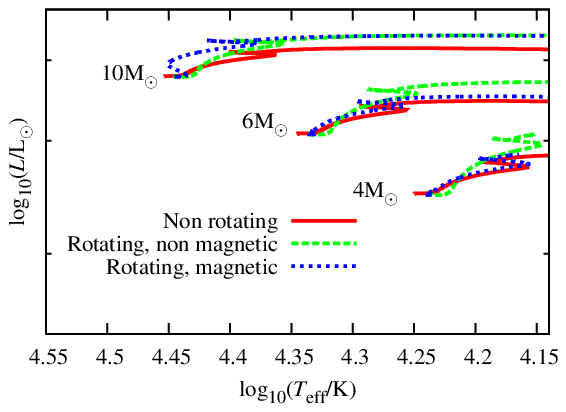}
\end{center}
\caption[Hertzsprung--Russell diagram for magnetic, rotating stars of various masses]{Hertzsprung--Russell diagram for stars with mass $4$\ms, $6$\ms\ and $10$\ms. The plot shows the predicted evolution for non--rotating stars, stars initially rotating at $300\,{\rm km\,s^{-1}}$ but with no magnetic field (c.f. case~1 from \citet{Potter11}) and magnetic stars initially rotating at $300\,{\rm \, km\,s^{-1}}$. In less massive stars magnetic braking rapidly spins down the star so the structural effects of rotation are much less apparent. In more massive stars the effect of braking is much weaker and so the evolution is much closer to the rotating, non--magnetic model.}
\label{fig.hr}
\end{figure}

Because less massive stars have stronger fields, both magnetically induced mixing and magnetic braking are much more effective in these stars. Owing to the stronger magnetic mixing, chemical transport is more efficient in less massive stars as discussed in section~\ref{sec.hunter}. As a result, more hydrogen is mixed down into the core of less massive stars. However, because magnetic braking causes lower--mass stars to spin down very rapidly, the effects on brightness and temperature that arise from changes in the stellar structure in rotating stars are far less apparent when magnetic fields are introduced, as shown in Fig.~\ref{fig.hr}. In the $10$\ms\ model we see that the difference between the magnetic and non--magnetic rotating models is smaller owing to the much weaker field and hence less rapid spin down. However, in the evolution of the $4$ and $6$\ms\ models the magnetic stars remain barely distinguishable from the evolution of the non--rotating stars.

\subsection{The lifetime of fossil fields}
\label{sec.fossilfield}

\begin{figure}
\begin{center}
\hspace{-0.5cm}
\includegraphics[width=0.49\textwidth]{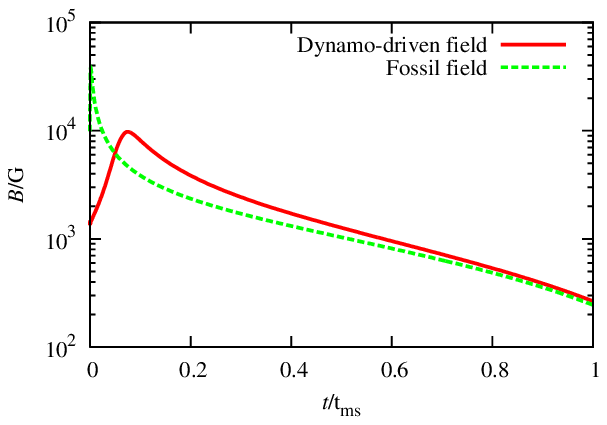}
\end{center}
\caption[Evolution of the magnetic field strength of $5$\ms\ in the absence of dynamo regeneration]{Evolution of two $5$\ms\ stars with different magnetic field models. The first star has no magnetic dynamo ($\gamma=0$) but starts with a very strong initial field ($B=10\,{\rm kG}$). The second star uses the same dynamo model and parameters as described in section~\ref{sec.model}. The star with an active dynamo is able to sustain the field for longer than the star with a fossil field but, owing to magnetic braking, both fields eventually decay exponentially. The two stars have similar field strengths at the end of the main sequence.}
\label{fig.fossil}
\end{figure}

An alternative to the radiative--dynamo model is that the magnetic field originates in the material that formed the star. If the protostellar cloud which forms a star is weakly magnetic, conservation of magnetic energy would result in a very strong main--sequence field. We call these fossil fields \citep{Braithwaite04}. In order for the fossil field model to work, the field must be able to survive the collapse of the protostellar cloud during the star formation process. The fossil field argument also relies on a stable field configuration being reached that would avoid destruction on main--sequence lifetimes. Certain stable configurations have been found in recent years \citep{Braithwaite06} and simulations have suggested that arbitrary field configurations do relax to these stable states \citep{Mathis11}. However, simple field configurations are still subject to the same instabilities as the fields we have generated by dynamo action, in particular the Tayler instability \citep{Tayler73}. There are a number of other instabilities that could occur in simple field configurations \citep{Parker66} but for now we consider only the Tayler instability.

We consider two stars, both initially rotating at $300\,{\rm km\,s^{-1}}$. The first star starts on the ZAMS with a magnetic field of $10\,{\rm kG}$ but $\gamma=0$ so no dynamo operates. The second star is a rotating magnetic star with dynamo parameters described in section~\ref{sec.model}. The evolution of the magnetic fields is shown in Fig.~\ref{fig.fossil}. In each case, the initial field undergoes some amplification at first owing to the onset of mass and angular momentum loss and the subsequent redistribution of angular momentum through the envelope. This is much more rapid in the case of the fossil field and does not appear in Fig.~\ref{fig.fossil}. The field then decays exponentially during the main sequence. We note that although the star with an operating dynamo is able to prevent the field from decaying for a short time, once magnetic braking has spun the star down sufficiently, the dynamo can no longer maintain the field which then decays exponentially. The final field strength is similar in each case.

Because the fossil field model predicts field evolution similar to that of the dynamo model it is difficult to argue which model is more physically accurate. However, we note that the fossil field strength has to be several orders of magnitude larger than the initial field in the case of a magnetic dynamo in order to reproduce the same final field. The question remains whether the fossil field argument can produce stars with strong enough initial fields so that they remain strong enough to influence chemical mixing in the star during the main sequence. \citet{Moss03} examined how much magnetic flux could potentially survive to the ZAMS from the pre--main sequence. He found that a significant fraction of flux could survive but only if the magnetic diffusivity was sufficiently low. Above this limit, no flux was expected to survive. The fossil field must also reproduce the two distinct observed populations in the Hunter diagram, shown in Fig.~\ref{fig.dufton}, discussed further in section~\ref{sec.hunter}. One could argue that this depends on the distribution of magnetic field strengths in protostellar clouds but the fossil field model must then also explain the mass--dependent distribution of field strengths observed in massive stars. Thus far we have come across no arguments that accurately reproduce these features of observed populations for fossil fields.

\subsection{Effect on surface composition}
\label{sec.hunter}

\begin{figure}
\begin{center}
\includegraphics[width=0.49\textwidth]{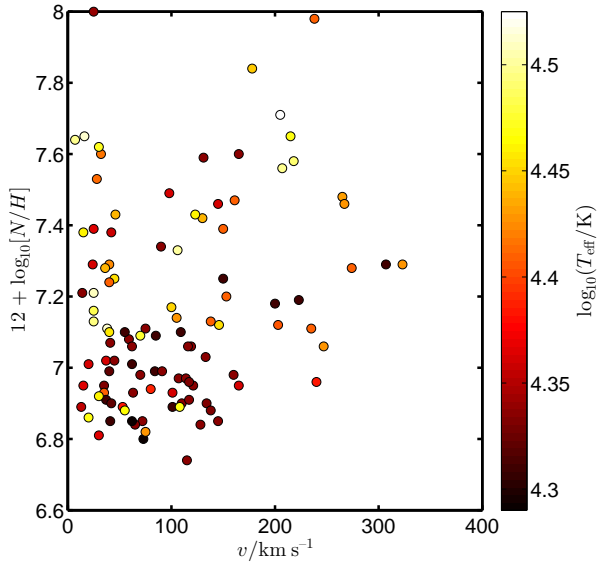} 
\end{center}
\caption[Hunter diagram for LMC stars observed in the VLT--FLAMES survey of massive stars]{Hunter diagram for the LMC stars observed in the VLT--FLAMES survey of massive stars \citep{Hunter09}. Stars with surface gravity smaller than $\log_{10}(g_{\rm eff}/{\rm cm^2\,s^{-1}})=3.2$ are classified as giants and have been excluded. The effective temperature of each star is also shown.}
\label{fig.dufton}
\end{figure}

\begin{figure}
\begin{center}
\includegraphics[width=0.49\textwidth]{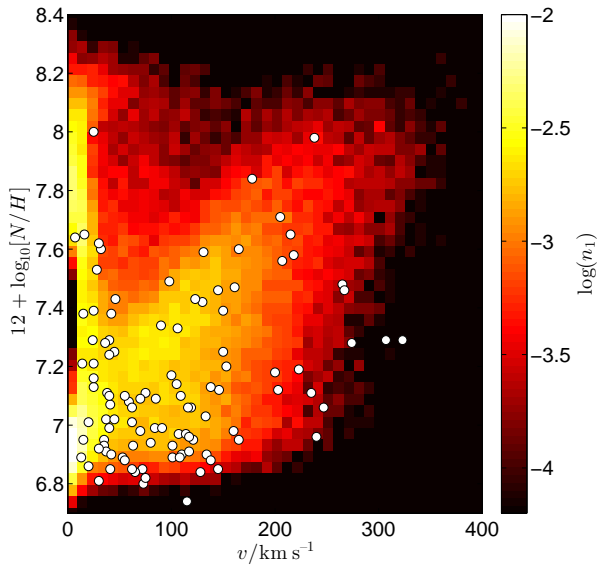} 
\end{center}
\caption[Hunter diagram for a population of stars drawn from the grid of models shown in Fig.~\ref{fig.grid}]{Hunter diagram for a population of stars drawn from the grid of models shown in Fig. \ref{fig.grid}. The population undergoes continuous star formation, is drawn from a Salpeter IMF and the velocity distribution is Gaussian with mean $\mu=145\,{\rm km\,s^{-1}}$ and standard deviation $\sigma=94\,{\rm km\,s^{-1}}$. The number of stars in each bin as a fraction of the total number of stars is $n_1$. The magnetic model reproduces well the two distinct populations of stars observed in the VLT--FLAMES survey. More massive stars which cannot support a dynamo are enriched by rotational mixing whereas lower--mass stars are spun down rapidly and are enriched by magnetic mixing.}
\label{fig.hunter}
\end{figure}

\begin{figure}
\begin{center}
\includegraphics[width=0.49\textwidth]{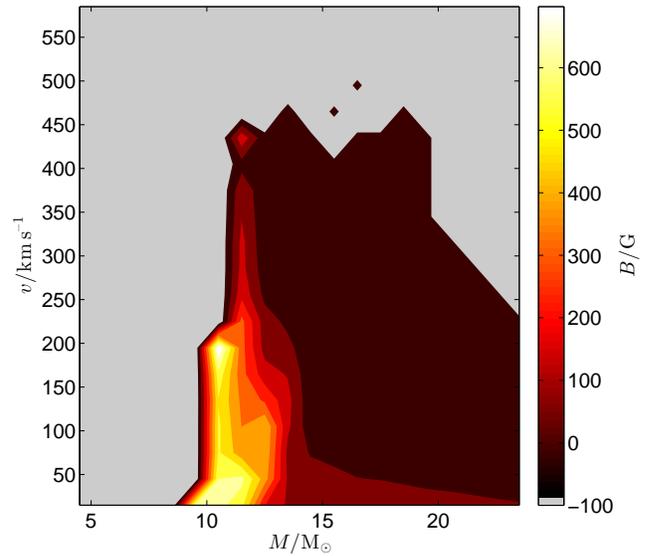} 
\end{center}
\caption[Relation between field strength, mass and rotation rate in a simulated population of stars according to the observational constraints of the VLT--FLAMES survey]{Distribution of magnetic field strengths with respect to mass and rotation rate for a population of stars undergoing continuous star formation. The population is drawn from a Salpeter IMF and the velocity distribution is Gaussian with mean $\mu=145\,{\rm km\,s^{-1}}$ and standard deviation $\sigma=94\,{\rm km\,s^{-1}}$. The gray region is where stars are not observed in the simulated population. Less massive stars are eliminated from the sample because they have insufficient magnitude for detection. The black region is for stars that appear in the simulated population but have no discernible field. We see that the magnetic stars in the sample, responsible for producing the slowly--rotating, enriched stars in Fig.~\ref{fig.hunter} come from a narrow region around $M=10$\ms.}
\label{fig.mvb2}
\end{figure}

The Hunter diagram \citep{Hunter09} is a plot of the surface nitrogen abundance in a star against surface velocity. The VLT--FLAMES survey of massive stars \citep{Evans05,Evans06,Dufton06} resulted in a significant amount of data on the nitrogen abundances in rotating stars in a number of samples from the Milky Way and Magellanic Clouds \citep{Hunter09}. In particular it was observed that there exists a class of stars that are slowly rotating ($v<60\,{\rm km\,s^{-1}}$) but exhibit significant nitrogen enrichment. It was suggested that these stars are, or once were, magnetic stars. If we extend the Hunter diagram to consider the effective temperature of each star as shown in Fig.~\ref{fig.dufton} we do not see a significant temperature variation between the two groups but we note that the mass range of stars in this sample is only $8<M/{\rm M_{\odot}}<20$ and so we cannot draw any strong conclusions about the relative mass distribution of the two enriched populations.

The observed distribution of surface abundance anomalies are well reproduced by our model which predicts magnetic fields only in stars less massive than around $15$\ms. The rest of the stars in the sample continue to evolve as non--magnetic stars as described in \citet{Potter12}. The two distinct populations that we see in Fig.~\ref{fig.dufton} are reflected by the predictions made in section~\ref{sec.mvb}, shown in Fig.~\ref{fig.hunter}. This shows a simulated population of stars between $8$\ms\ and $20$\ms\ with our radiative--dynamo model and with magnetic braking. The mass range is smaller than that of the full grid shown in Fig.~\ref{fig.grid} owing to the removal of the least and most massive stars because of observational effects. The population was generated with the population synthesis code {\sc starmaker} \citep{Brott11}. It behaves exactly as we would expect from the VLT--FLAMES data. The stars initially have a full spread of rotation rates but the magnetic population spins down rapidly owing to the effects of magnetic braking. The magnetic field continues to affect the mixing and these stars become enriched as they age producing a population of magnetic, slowly rotating, chemically peculiar stars. More massive stars, where an equilibrium field cannot be supported by the dynamo, evolve as non--magnetic stars with hydrodynamic turbulence driving the mixing. This produces a second population whose enrichment increases with rotation rate as modelled in \citet{Potter12}. The two populations are also highlighted in Fig.~\ref{fig.mvb2} which shows the relationship between field strength, mass and rotation rate in the simulated population. Most of the stars evolve without magnetic fields but there is a small region, at the lower mass limit of the sample ($M\approx 10$\ms), where stars are predicted to be magnetic. We note that, because this region is very narrow, small changes in the boundary between magnetic and non--magnetic evolution have a significant effect on the number of magnetic stars in the observed sample. It is possible this effect could be produced by fossil fields as discussed in section~\ref{sec.fossilfield} but thus far there is no way to explain why we see such distinct populations in the VLT--FLAMES data or why magnetic fields have a higher incidence rate amongst less massive stars.

We also note that those stars in Fig.~\ref{fig.dufton} with nitrogen enrichment $6.8<\log_{10}[N/H]<7.1$ and $0<v/{\rm km\,s^{-1}}<150$ cannot easily be categorized into either group of stars. They may be low--mass, fast rotators that have been partially spun down by magnetic braking, low--mass stars that are born with slow rotation or high--mass stars that are born with slow rotation. These stars evolve along a relatively similar path in the Hunter diagram.

\subsection{Variation with different parameters}
\label{sec.calibration}

\begin{table*}
\begin{center}
\begin{tabular}{cccccc}
\hline
$C_{\rm m}$&${\rm Pr_m}$&$\gamma$&$\max\left(\frac{rB_{\phi}}{A}\right)$&$\max(q)$&$\max(B_{\rm surf}/{\rm G})$\\
\hline
$1$&$1$&$1.93\times 10^{-16}$&$4.2\times 10^9$&$1.39$&$3.06\times 10^3$\\
$0.1$&$1$&$2.40\times 10^{-18}$&$1.1\times 10^{11}$&$0.96$&$7.10\times 10^2$\\
$10$&$1$&$3.53\times 10^{-14}$&$1.2\times 10^{8}$&$0.019$&$1.53\times 10^3$\\
$1$&$0.1$&$1.52\times 10^{-16}$&$3.74\times 10^{9}$&$1.04$&$7.40\times 10^{2{\rm (2)}}$\\
$0.1$&$0.1$&$3.00\times 10^{-17{\rm (3)}}$&&&\\
$10$&$0.1$&$1.81\times 10^{-14}$&$4.4\times 10^{8}$&$0.25$&$1.27\times 10^3$\\
$1$&$10$&$1.04\times 10^{-14}$&$8.13\times 10^{8}$&$0.0089$&$1.55\times 10^3$\\
$0.1$&$10$&$4.47\times 10^{-16}$&$1.95\times 10^{10}$&$0.14$&$8.92\times 10^2$\\
$10$&$10$&$1.58\times 10^{-11}$&$2.1\times 10^{7}$&$0.033$&${\rm N/a}^{(1)}$\\
\hline
\end{tabular}
\caption[The variation of magnetic stellar evolution owing to variation in parameters for magnetic field evolution]{The variation of magnetic and stellar parameters with different values for $C_{\rm m}$ and ${\rm Pr_m}$ for a $5$\ms\ star initially rotating at $200\,{\rm km\,s^{-1}}$ with magnetic braking. Each model was taken to have the same equilibrium ZAMS field. The table shows the values of $C_{\rm m}$, ${\rm Pr_m}$ and $\gamma$ used for each model as well as the maximum internal value of the ratio of the poloidal and toroidal field, $rB_{\phi}/A$ and $q=\partial(\log\Omega)/\partial(\log r)$, taken at $5 \times 10^7 {\rm yr}$. Finally the table shows the maximum value of the surface field during the main sequence. We note three special entries in the table. (1) This star evolved quasi--homogeneously and produced a monotonically increasing field well beyond the normal main--sequence lifetime. Therefore defining a maximum main--sequence field was inappropriate. (2) This star evolved normally but we note that for a slightly smaller value of $\gamma$ we were unable to maintain an equilibrium field. This effect was discussed in section \ref{sec.mvb}. (3) This star is similar to (2) but in this case we were totally unable to maintain an equilibrium field at the desired strength. We note that for stars (2) and (3), a stronger field can be maintained provided the dynamo--efficiency is sufficiently large.}
\label{tab.calibration}
\end{center}
\end{table*}

The model currently contains four parameters which we may vary independently. If we include possible recalibration of the Alfv\'{e}n radius by constants of order unity then this increases to five. We may fix the Alfv\'{e}n radius by ensuring that the population of enriched magnetic stars is confined to the appropriate band of rotation rates as discussed in section~\ref{sec.hunter}. We can also set ${\rm Pr_c}$ by ensuring that the maximal enrichment of magnetic stars is the same as in the VLT--FLAMES data also discussed in section~\ref{sec.hunter}. The remaining three parameters may then be varied so that typical field strengths are of the order $10\,{\rm kG}$, as observed in magnetic Ap stars \citep{Mathys09}. This value is subject to change though given the scarcity of observations of magnetic stars. This still leaves a high degree of freedom within the model. Up to this point we have used $C={\rm Pr_{\rm m}}=1$ and $\gamma=10^{-15}$ but we consider the effect of varying $C_{\rm m}$ and ${\rm Pr_m}$ by an order of magnitude in either direction. We ran our $5$\ms\ star initially rotating at $300\,{\rm \, km\,s^{-1}}$ with magnetic braking. The effect on a number of parameters is shown in Table~\ref{tab.calibration}.

For low magnetic Prandtl numbers it is much more difficult to sustain the dynamo. The same surface field is reproduced with smaller dynamo efficiencies but the minimum sustainable field strength is larger. In the case of small $C_{\rm m}$ and small ${\rm Pr_m}$, the field was completely quenched by the hydrodynamic turbulence as described in section \ref{sec.mvb}. A dynamo could be sustained for stronger surface fields but only by increasing the dynamo efficiency significantly. Even for $C_{\rm m}=1$ we found that for a small reduction in $\gamma$ the ZAMS field collapsed.

For higher values of $C_{\rm m}$, the diffusion of the magnetic field requires a larger dynamo efficiency in order to maintain the same strength field and vice--versa for smaller values of $C_{\rm m}$. For simultaneously large values of $C_{\rm m}$ and ${\rm Pr_m}$ the field keeps growing monotonically with time during quasi--homogeneous evolution. This is to be expected when the dynamo--driven mixing becomes very high. Typically we could adjust $\rm Pr_c$ to compensate.

Regardless of our choice of $C_{\rm m}$ and ${\rm Pr_m}$, the ratio of the poloidal and toroidal field strength is well correlated with the dynamo efficiency. Larger values of the dynamo efficiency lead to a smaller ratio between the two field strengths. This is because of the form of equations (\ref{eq.toroidalfield}) and (\ref{eq.poloidalfield}). Because the two fields have the same diffusion timescales, their equilibria depend on the regeneration terms. In the case of the poloidal field this comes from the $\alpha$--effect and for the toroidal field it comes from the shear. In all of our models, the $\alpha$--effect is much weaker than the effect of shear and so the poloidal field strength is much smaller. However if $\gamma$ is increased, increasing the regeneration of the poloidal field but having little direct effect on the toroidal field, the ratio between the two becomes much smaller.

There are other aspects of the evolution that are much more difficult to explain and are related to the non--linearities in the model and their coupling to the effects of stellar evolution on nuclear timescales. We might expect the maximum value of the shear to always be smaller with higher values of $C_{\rm m}$ because the angular momentum transport is more efficient but, while this is true in general, it isn't a simple relationship. Likewise the maximum main--sequence surface field doesn't seem to correlate with either free parameter. 

In particular, the relative abundance of slow and fast rotating chemically peculiar stars may be explained by a shift in the position by mass of the cut--off between magnetic stars and non--magnetic stars discussed in section \ref{sec.mvb}. The effect of these free parameters on the position of the cut--off is something we leave for future work.

\section{Discussion}
\label{sec.discussion}

Magnetic fields are one of the most mysterious and least understood aspects of stellar evolution. The first magnetic massive star was discovered over 65 years ago \citep{Babcock47} and yet debate still rages about whether these fields have primordial origin or are generated by a radiative dynamo acting within the stellar envelope. Models of magnetic stars must reproduce the observed phenomenon of magnetic A stars with unusual surface compositions that have much slower rotation rates than the rest of their  population \citep{Mathys04}. The data from the VLT--FLAMES survey of massive stars \citep{Evans05, Evans06} also supports the idea that there exists a population of stars that are slowly rotating but have a high degree of nitrogen enrichment \citep{Hunter09}.

We have presented a simple radiative dynamo model that arises because of the Tayler pinch--type instability \citep{Tayler73} and is based on the model of \citet{Spruit99} which was further developed by \citet{Spruit02} and \citet{Maeder04}. Unlike previous work, we have evolved both the poloidal and toroidal fields as independent variables at each radius in the star coupled to the angular momentum distribution of the star. The magnetic fields evolve according to a latitudinally--averaged induction equation with the inclusion of an $\alpha\Omega$--dynamo mechanism derived from mean--field magnetohydrodynamics \citep{Schmalz91}. We introduce a model for magnetic braking similar to that of \citet{ud-Doula02}. The model depends on a number of parameters, the overall strength of the magnetic turbulence, the magnetic Prandtl number, the chemical Prandtl number, the dynamo efficiency and the critical ratio of the kinetic energy to the magnetic energy, which defines the Alfv\'{e}n radius. The choices of $C_{\rm m}$, which affects the strength of the magnetic turbulence, and ${\rm Pr_c}$ have a strong effect on the dynamo efficiency needed to sustain the field but the relation between these parameters and the internal evolution of the models is complicated.

In models of the magnetic field, when magnetic braking is not included, the field varies only by a factor of a few during the main sequence. When we include magnetic braking, the Alfv\'{e}n radius is typically between $10$ and $100$ times greater than the stellar radius and so angular momentum loss is some $10^3$ times greater than from non--magnetic mass loss alone. The rapid angular momentum loss from the surface drives additional shear that leads to increased field generation. In magnetic stars with magnetic braking, the field increases rapidly at the start of the main sequence before decaying exponentially. The field strengths at the end of the main sequence are predicted to be of order $100{\rm \, G}$.

We consider a population of stars with this magnetic model and find two distinct types of behaviour. For stars more massive than around $15$\ms\ the Kelvin--Helmholtz turbulence dominates over the magnetic turbulence and a stable field cannot be sustained by the dynamo. In these cases we see no appreciable field strength during the main sequence so the stars evolve according to our normal prescription for non--magnetic, rotating stars. The predicted field strength is stronger for rapid rotators but the overall strength does not depend strongly on the stellar mass except near the limit at which the dynamo can sustain the field. Although the magnetic field decays exponentially after an initial peak, it remains strong enough to have a significant effect on the chemical evolution of the star. Though the actual mass at which this dichotomy sets in depends on parameters, the fact it exists is an important consequence of our model.

If we look at the evolution of an artificially strong initial field in the absence of any dynamo action, but subject to the diffusion that arises from the \citet{Tayler73} instability, we find that reproducing the same TAMS field requires an initial field several orders of magnitude larger than in the presence of a dynamo because any fossil field is predicted to decay exponentially. The fossil field hypothesis suffers from the problem that we expect the fields in low--mass stars to decay more than in more massive stars, likely because of their much longer main--sequence lifetimes. This is opposite to observed trends which suggest that less massive stars are more likely to support strong fields than more massive stars \citet{Grunhut11}. This model also offers no explanation as to why we see two distinct groups in the Hunter diagram. Both of these issues are well resolved by our $\alpha\Omega$--dynamo model.

We created an artificial population of stars with the population synthesis code {\sc starmaker} \citep{Brott11}, including the effects of the ${\alpha\Omega}$--dynamo and magnetic braking. The population reflects well the observations of the VLT--FLAMES survey of massive stars. The survey observed two distinct populations of stars. The first shows increasing nitrogen enrichment with rotation rate, the second is a class of slow--rotating stars that exhibit unusually high nitrogen abundances compared to the rest of the population. This distribution of stars is well reproduced by the magnetic model. The fact that the two very different evolutionary paths arise naturally from the model is very encouraging to explain why we observe these two classes of star without having to appeal to the fossil fields argument.

There are still a number of open questions and further refinements that need to be made to the model. We have evolved a magnetic population of stars with the same initial velocity distribution as the non--magnetic stars. If the radiative dynamo has a strong effect on the pre--mainsequence evolution then magnetic braking causes magnetic stars to reach the ZAMS with significantly slower rotation rates than stars with no significant field. This is indeed observed in stellar populations \citep{Mathys04}. \citet{Alecian08} also discovered a number of stars on the pre--main sequence which exhibited significant magnetic activity. They attribute these to fossil fields by eliminating the possibility that the fields could be generated by a convective dynamo. However, if a radiative dynamo operates in these stars it could also be responsible for the generation of the observed fields. By comparison, the observations of \citet{Grunhut12} suggest that magnetic stars may reach the main sequence with significant rotational velocities. If magnetic stars were born with slower rotation rates than their non--magnetic counterparts then this would partly explain why the required dynamo efficiency is so small and why the predicted ratio between the poloidal and toroidal fields is so large. If magnetic stars were born with lower surface rotation rates then a higher dynamo efficiency would be needed to produce observed magnetic field strengths. This would reduce the difference between the $\alpha$--effect and the $\Omega$--effect and so the ratio of the strengths of the poloidal and toroidal fields would be closer to unity. Another possible explanation for why the predicted dynamo efficiency is so small is that we chose the radial coordinate as the length scale for the dynamo action. In reality a more sensible choice may have been the length scale of the saturated magnetic instabilities, $l_r$ (c.f. equations~(\ref{eq.lr}) and~(\ref{eq.lr2})). A shorter length scale would result in a weaker dynamo and therefore the dynamo efficiency would need to be higher to sustain the same field strength.

The observed proportion of Ap stars as a fraction of the whole population of A stars is roughly $10\%$ \citep{Moss01}. Our grid of models does not yet extend down to the mass range for A stars ($1.4<M/$\ms$<2.1$) and so we cannot yet say whether our population matches this statistic. We do expect that, given the predicted initial velocity distribution of massive stars, the population of A stars should still be dominated by slow rotators that do not support a radiative dynamo. In the mass range of our simulated population, over $90$\% of stars in the sample have surface field strength less than $187$\,G. This is well below the limit of $300$\,G anticipated for the transition to Ap classification \citep{Auriere07}. Although our population contains some very massive stars where we expect smaller field strengths, the form of the IMF ensures that the population is still dominated by intermediate--mass stars so the figure for A stars is likely to be similar. It is also likely that below a certain amount of shear, a dynamo does not operate. We have not taken this into account in our simple model. If it is the case, there may also be a sharp transition between magnetic and non--magnetic behaviour at low rotation rates.

In our models we have assumed a simple magnetic field geometry. Even if real fields are generated by dynamo action then they may still relax to stable field configurations such as those suggested by \citet{Braithwaite06,Mathis11}. Further work is needed to determine how the model might behave differently under these conditions. Further consideration must also be given to the action of convection on the magnetic field. Does our diffusive model apply in convective zones and if so is it anisotropic? Furthermore, can we better constrain the free parameters in the system, including the efficiency of magnetic braking? Although data on magnetic stars is scarce, a great deal of progress has been made possible by surveys such as the VLT--FLAMES survey of massive stars and the MiMeS project. These provide sufficient clues to further constrain our existing models. Additional progress will no doubt be possible thanks to ongoing developments in stellar observations from the MiMeS project \citep{Wade09, Grunhut11} and additional data on stellar surface compositions through projects such as the VLT--FLAMES tarantula survey \citep{Evans11}.

\section{Conclusions}
\label{sec.conclusions}

We have presented a new model for a radiative $\alpha$--$\Omega$ dynamo. The model is based on the Tayler--Spruit dynamo \citep{Spruit02} and incorporates magnetic braking based on \citet{ud-Doula02}. The model predicts two distinct populations of massive stars. 

\begin{itemize}
\item In stars with masses greater than around $15$\ms, the dynamo cannot be sustained. These stars evolve as described in \citet{Potter11}; purely hydrodynamic turbulence caused by rotation results in surface abundance anomalies. The degree of chemical peculiarity is correlated strongly with the rotation rate. Less massive stars with low rotation rates also evolve this way. 

\item Less massive stars with sufficiently high rotation rates have an active dynamo and so exhibit strong magnetic fields. These stars are spun down quickly by magnetic braking and turbulence causes changes to the surface composition. These stars appear as slowly--rotating chemically--peculiar stars in the Hunter diagram \citep{Hunter09}.
\end{itemize}

These two populations were observed in the VLT--FLAMES survey of massive stars \citep{Hunter09} and the predicted number of magnetic stars in our simulated sample matches well with observations \citep{Moss01}. Further data is needed to confirm the mass--dependency of the simulated field strengths \citep{Grunhut12}. Although this model does not rule out the possibility that magnetic fields in massive stars have a fossil--field origin, it does strongly suggest that they may instead be the result of a radiative $\alpha$--$\Omega$ dynamo.

\section{Acknowledgements}
ATP thanks I. Brott for allowing use of the \textsc{starmaker} code and the STFC for his studentship. SMC is grateful to the Cambridge Institute of Astronomy for supporting his summer visit. CAT thanks Churchill college for his fellowship.

\bibliographystyle{mn2e}
\bibliography{paper7}

\begin{thebibliography}{57}
\expandafter\ifx\csname natexlab\endcsname\relax\def\natexlab#1{#1}\fi

\bibitem[{{Alecian} {et~al.}(2008){Alecian}, {Wade}, {Catala}, {Folsom},
  {Grunhut}, {Donati}, {Petit}, {Bagnulo}, {Marsden}, {Ramirez Velez},
  {Landstreet}, {Boehm}, {Bouret}, \& {Silvester}}]{Alecian08}
{Alecian} E., {Wade} G.~A., {Catala} C., {Folsom} C., {Grunhut} J., {Donati}
  J.-F., {Petit} P., {Bagnulo} S., {Marsden} S.~C., {Ramirez Velez} J.~C.,
  {Landstreet} J.~D., {Boehm} T., {Bouret} J.-C., {Silvester} J., 2008,
  Contributions of the Astronomical Observatory Skalnate Pleso, 38, 235

\bibitem[{{Auri{\`e}re} {et~al.}(2007){Auri{\`e}re}, {Wade}, {Silvester},
  {Ligni{\`e}res}, {Bagnulo}, {Bale}, {Dintrans}, {Donati}, {Folsom},
  {Gruberbauer}, {Hui Bon Hoa}, {Jeffers}, {Johnson}, {Landstreet},
  {L{\`e}bre}, {Lueftinger}, {Marsden}, {Mouillet}, {Naseri}, {Paletou},
  {Petit}, {Power}, {Rincon}, {Strasser}, \& {Toqu{\'e}}}]{Auriere07}
{Auri{\`e}re} M., {Wade} G.~A., {Silvester} J., {Ligni{\`e}res} F., {Bagnulo}
  S., {Bale} K., {Dintrans} B., {Donati} J.~F., {Folsom} C.~P., {Gruberbauer}
  M., {Hui Bon Hoa} A., {Jeffers} S., {Johnson} N., {Landstreet} J.~D.,
  {L{\`e}bre} A., {Lueftinger} T., {Marsden} S., {Mouillet} D., {Naseri} S.,
  {Paletou} F., {Petit} P., {Power} J., {Rincon} F., {Strasser} S., {Toqu{\'e}}
  N., 2007, \aap, 475, 1053

\bibitem[{{Babcock}(1947)}]{Babcock47}
{Babcock} H.~W., 1947, \apj, 105, 105

\bibitem[{{Bagnulo} {et~al.}(2004){Bagnulo}, {Hensberge}, {Landstreet},
  {Szeifert}, \& {Wade}}]{Bagnulo04}
{Bagnulo} S., {Hensberge} H., {Landstreet} J.~D., {Szeifert} T., {Wade} G.~A.,
  2004, \aap, 416, 1149

\bibitem[{{B{\"o}hm-Vitense}(1958)}]{Bohm58}
{B{\"o}hm-Vitense} E., 1958, \zap, 46, 108

\bibitem[{{Borra} \& {Landstreet}(1978)}]{Borra78}
{Borra} E.~F., {Landstreet} J.~D., 1978, \apj, 222, 226

\bibitem[{{Braithwaite} \& {Nordlund}(2006)}]{Braithwaite06}
{Braithwaite} J., {Nordlund} {\AA}., 2006, \aap, 450, 1077

\bibitem[{{Braithwaite} \& {Spruit}(2004)}]{Braithwaite04}
{Braithwaite} J., {Spruit} H.~C., 2004, \nat, 431, 819

\bibitem[{{Brandenburg}(2001)}]{Brandenburg01}
{Brandenburg} A., 2001, \apj, 550, 824

\bibitem[{{Brott} {et~al.}(2011{\natexlab{a}}){Brott}, {de Mink}, {Cantiello},
  {Langer}, {de Koter}, {Evans}, {Hunter}, {Trundle}, \& {Vink}}]{Brott11b}
{Brott} I., {de Mink} S.~E., {Cantiello} M., {Langer} N., {de Koter} A.,
  {Evans} C.~J., {Hunter} I., {Trundle} C., {Vink} J.~S., 2011{\natexlab{a}},
  \aap, 530, A115

\bibitem[{{Brott} {et~al.}(2011{\natexlab{b}}){Brott}, {Evans}, {Hunter}, {de
  Koter}, {Langer}, {Dufton}, {Cantiello}, {Trundle}, {Lennon}, {de Mink},
  {Yoon}, \& {Anders}}]{Brott11}
{Brott} I., {Evans} C.~J., {Hunter} I., {de Koter} A., {Langer} N., {Dufton}
  P.~L., {Cantiello} M., {Trundle} C., {Lennon} D.~J., {de Mink} S.~E., {Yoon}
  S.-C., {Anders} P., 2011{\natexlab{b}}, \aap, 530, A116

\bibitem[{{Cowling}(1945)}]{Cowling45}
{Cowling} T.~G., 1945, \mnras, 105, 166

\bibitem[{{Donati} {et~al.}(2002){Donati}, {Babel}, {Harries}, {Howarth},
  {Petit}, \& {Semel}}]{Donati02}
{Donati} J.-F., {Babel} J., {Harries} T.~J., {Howarth} I.~D., {Petit} P.,
  {Semel} M., 2002, \mnras, 333, 55

\bibitem[{{Donati} {et~al.}(2006{\natexlab{a}}){Donati}, {Howarth}, {Bouret},
  {Petit}, {Catala}, \& {Landstreet}}]{Donati06}
{Donati} J.-F., {Howarth} I.~D., {Bouret} J.-C., {Petit} P., {Catala} C.,
  {Landstreet} J., 2006{\natexlab{a}}, \mnras, 365, L6

\bibitem[{{Donati} {et~al.}(2006{\natexlab{b}}){Donati}, {Howarth}, {Jardine},
  {Petit}, {Catala}, {Landstreet}, {Bouret}, {Alecian}, {Barnes}, {Forveille},
  {Paletou}, \& {Manset}}]{Donati06b}
{Donati} J.-F., {Howarth} I.~D., {Jardine} M.~M., {Petit} P., {Catala} C.,
  {Landstreet} J.~D., {Bouret} J.-C., {Alecian} E., {Barnes} J.~R., {Forveille}
  T., {Paletou} F., {Manset} N., 2006{\natexlab{b}}, \mnras, 370, 629

\bibitem[{{Donati} {et~al.}(2001){Donati}, {Wade}, {Babel}, {Henrichs}, {de
  Jong}, \& {Harries}}]{Donati01}
{Donati} J.-F., {Wade} G.~A., {Babel} J., {Henrichs} H.~f., {de Jong} J.~A.,
  {Harries} T.~J., 2001, \mnras, 326, 1265

\bibitem[{{Dufton} {et~al.}(2006){Dufton}, {Smartt}, {Lee}, {Ryans}, {Hunter},
  {Evans}, {Herrero}, {Trundle}, {Lennon}, {Irwin}, \& {Kaufer}}]{Dufton06}
{Dufton} P.~L., {Smartt} S.~J., {Lee} J.~K., {Ryans} R.~S.~I., {Hunter} I.,
  {Evans} C.~J., {Herrero} A., {Trundle} C., {Lennon} D.~J., {Irwin} M.~J.,
  {Kaufer} A., 2006, \aap, 457, 265

\bibitem[{{Eggleton}(1971)}]{Eggleton71}
{Eggleton} P.~P., 1971, \mnras, 151, 351

\bibitem[{{Evans} {et~al.}(2006){Evans}, {Lennon}, {Smartt}, \&
  {Trundle}}]{Evans06}
{Evans} C.~J., {Lennon} D.~J., {Smartt} S.~J., {Trundle} C., 2006, \aap, 456,
  623

\bibitem[{{Evans} {et~al.}(2005){Evans}, {Smartt}, {Lee}, {Lennon}, {Kaufer},
  {Dufton}, {Trundle}, {Herrero}, {Sim{\'o}n-D{\'{\i}}az}, {de Koter},
  {Hamann}, {Hendry}, {Hunter}, {Irwin}, {Korn}, {Kudritzki}, {Langer},
  {Mokiem}, {Najarro}, {Pauldrach}, {Przybilla}, {Puls}, {Ryans}, {Urbaneja},
  {Venn}, \& {Villamariz}}]{Evans05}
{Evans} C.~J., {Smartt} S.~J., {Lee} J.-K., {Lennon} D.~J., {Kaufer} A.,
  {Dufton} P.~L., {Trundle} C., {Herrero} A., {Sim{\'o}n-D{\'{\i}}az} S., {de
  Koter} A., {Hamann} W.-R., {Hendry} M.~A., {Hunter} I., {Irwin} M.~J., {Korn}
  A.~J., {Kudritzki} R.-P., {Langer} N., {Mokiem} M.~R., {Najarro} F.,
  {Pauldrach} A.~W.~A., {Przybilla} N., {Puls} J., {Ryans} R.~S.~I., {Urbaneja}
  M.~A., {Venn} K.~A., {Villamariz} M.~R., 2005, \aap, 437, 467

\bibitem[{{Evans} {et~al.}(2011){Evans}, {Taylor}, {H{\'e}nault-Brunet},
  {Sana}, {de Koter}, {Sim{\'o}n-D{\'{\i}}az}, {Carraro}, {Bagnoli}, {Bastian},
  {Bestenlehner}, {Bonanos}, {Bressert}, {Brott}, {Campbell}, {Cantiello},
  {Clark}, {Costa}, {Crowther}, {de Mink}, {Doran}, {Dufton}, {Dunstall},
  {Friedrich}, {Garcia}, {Gieles}, {Gr{\"a}fener}, {Herrero}, {Howarth},
  {Izzard}, {Langer}, {Lennon}, {Ma{\'{\i}}z Apell{\'a}niz}, {Markova},
  {Najarro}, {Puls}, {Ramirez}, {Sab{\'{\i}}n-Sanjuli{\'a}n}, {Smartt},
  {Stroud}, {van Loon}, {Vink}, \& {Walborn}}]{Evans11}
{Evans} C.~J., {Taylor} W.~D., {H{\'e}nault-Brunet} V., {Sana} H., {de Koter}
  A., {Sim{\'o}n-D{\'{\i}}az} S., {Carraro} G., {Bagnoli} T., {Bastian} N.,
  {Bestenlehner} J.~M., {Bonanos} A.~Z., {Bressert} E., {Brott} I., {Campbell}
  M.~A., {Cantiello} M., {Clark} J.~S., {Costa} E., {Crowther} P.~A., {de Mink}
  S.~E., {Doran} E., {Dufton} P.~L., {Dunstall} P.~R., {Friedrich} K., {Garcia}
  M., {Gieles} M., {Gr{\"a}fener} G., {Herrero} A., {Howarth} I.~D., {Izzard}
  R.~G., {Langer} N., {Lennon} D.~J., {Ma{\'{\i}}z Apell{\'a}niz} J., {Markova}
  N., {Najarro} F., {Puls} J., {Ramirez} O.~H., {Sab{\'{\i}}n-Sanjuli{\'a}n}
  C., {Smartt} S.~J., {Stroud} V.~E., {van Loon} J.~T., {Vink} J.~S., {Walborn}
  N.~R., 2011, \aap, 530, A108

\bibitem[{{Grunhut} {et~al.}(2012){Grunhut}, {Rivinius}, {Wade}, {Townsend},
  {Marcolino}, {Bohlender}, {Szeifert}, {Petit}, {Matthews}, {Rowe}, {Moffat},
  {Kallinger}, {Kuschnig}, {Guenther}, {Rucinski}, {Sasselov}, \&
  {Weiss}}]{Grunhut12}
{Grunhut} J.~H., {Rivinius} T., {Wade} G.~A., {Townsend} R.~H.~D., {Marcolino}
  W.~L.~F., {Bohlender} D.~A., {Szeifert} T., {Petit} V., {Matthews} J.~M.,
  {Rowe} J.~F., {Moffat} A.~F.~J., {Kallinger} T., {Kuschnig} R., {Guenther}
  D.~B., {Rucinski} S.~M., {Sasselov} D., {Weiss} W.~W., 2012, \mnras, 419,
  1610

\bibitem[{{Grunhut} {et~al.}(2011){Grunhut}, {Wade}, \& {the MiMeS
  Collaboration}}]{Grunhut11}
{Grunhut} J.~H., {Wade} G.~A., {the MiMeS Collaboration}, 2011, ArXiv e-prints

\bibitem[{{Heger} {et~al.}(2000){Heger}, {Langer}, \& {Woosley}}]{Heger00}
{Heger} A., {Langer} N., {Woosley} S.~E., 2000, \apj, 528, 368

\bibitem[{{Hubrig} {et~al.}(2005){Hubrig}, {Nesvacil}, {Sch{\"o}ller}, {North},
  {Mathys}, {Kurtz}, {Wolff}, {Szeifert}, {Cunha}, \& {Elkin}}]{Hubrig05}
{Hubrig} S., {Nesvacil} N., {Sch{\"o}ller} M., {North} P., {Mathys} G., {Kurtz}
  D.~W., {Wolff} B., {Szeifert} T., {Cunha} M.~S., {Elkin} V.~G., 2005, \aap,
  440, L37

\bibitem[{{Hunter} {et~al.}(2009){Hunter}, {Brott}, {Langer}, {Lennon},
  {Dufton}, {Howarth}, {Ryans}, {Trundle}, {Evans}, {de Koter}, \&
  {Smartt}}]{Hunter09}
{Hunter} I., {Brott} I., {Langer} N., {Lennon} D.~J., {Dufton} P.~L., {Howarth}
  I.~D., {Ryans} R.~S.~I., {Trundle} C., {Evans} C.~J., {de Koter} A., {Smartt}
  S.~J., 2009, \aap, 496, 841

\bibitem[{{Maeder}(2003)}]{Maeder03}
{Maeder} A., 2003, \aap, 399, 263

\bibitem[{{Maeder} \& {Meynet}(2000)}]{Maeder00}
{Maeder} A., {Meynet} G., 2000, \aap, 361, 159

\bibitem[{{Maeder} \& {Meynet}(2003)}]{Maeder03b}
---, 2003, \aap, 411, 543

\bibitem[{{Maeder} \& {Meynet}(2004)}]{Maeder04}
---, 2004, \aap, 422, 225

\bibitem[{{Mathis} {et~al.}(2011){Mathis}, {Duez}, \& {Braithwaite}}]{Mathis11}
{Mathis} S., {Duez} V., {Braithwaite} J., 2011, in IAU Symposium, Vol. 271, IAU
  Symposium, {N.~H.~Brummell, A.~S.~Brun, M.~S.~Miesch, \& Y.~Ponty}, ed., pp.
  270--278

\bibitem[{{Mathys}(2004)}]{Mathys04}
{Mathys} G., 2004, in IAU Symposium, Vol. 224, The A-Star Puzzle, {J.~Zverko,
  J.~Ziznovsky, S.~J.~Adelman, \& W.~W.~Weiss}, ed., pp. 225--234

\bibitem[{{Mathys}(2009)}]{Mathys09}
---, 2009, in Astronomical Society of the Pacific Conference Series, Vol. 405,
  Solar Polarization 5: In Honor of Jan Stenflo, {S.~V.~Berdyugina,
  K.~N.~Nagendra, \& R.~Ramelli}, ed., p. 473

\bibitem[{{Meynet} \& {Maeder}(2000)}]{Meynet00}
{Meynet} G., {Maeder} A., 2000, \aap, 361, 101

\bibitem[{{Moss}(2001)}]{Moss01}
{Moss} D., 2001, in Astronomical Society of the Pacific Conference Series, Vol.
  248, Magnetic Fields Across the Hertzsprung-Russell Diagram, {G.~Mathys,
  S.~K.~Solanki, \& D.~T.~Wickramasinghe}, ed., p. 305

\bibitem[{{Moss}(2003)}]{Moss03}
---, 2003, \aap, 403, 693

\bibitem[{{Neiner} {et~al.}(2003){Neiner}, {Henrichs}, {Floquet}, {Fr{\'e}mat},
  {Preuss}, {Hubert}, {Geers}, {Tijani}, {Nichols}, \& {Jankov}}]{Neiner03}
{Neiner} C., {Henrichs} H.~F., {Floquet} M., {Fr{\'e}mat} Y., {Preuss} O.,
  {Hubert} A.-M., {Geers} V.~C., {Tijani} A.~H., {Nichols} J.~S., {Jankov} S.,
  2003, \aap, 411, 565

\bibitem[{{Nordhaus}(2010)}]{Nordhaus10}
{Nordhaus} J., 2010, in Astronomical Society of the Pacific Conference Series,
  Vol. 432, New Horizons in Astronomy: Frank N. Bash Symposium 2009,
  {L.~M.~Stanford, J.~D.~Green, L.~Hao, \& Y.~Mao}, ed., p. 117

\bibitem[{{Parker}(1958)}]{Parker58}
{Parker} E.~N., 1958, \apj, 128, 664

\bibitem[{{Parker}(1966)}]{Parker66}
---, 1966, \apj, 145, 811

\bibitem[{{Petit} \& {Wade}(2011)}]{Petit11}
{Petit} V., {Wade} G.~A., 2011, ArXiv e-prints

\bibitem[{{Pols} {et~al.}(1995){Pols}, {Tout}, {Eggleton}, \& {Han}}]{Pols95}
{Pols} O.~R., {Tout} C.~A., {Eggleton} P.~P., {Han} Z., 1995, \mnras, 274, 964

\bibitem[{{Potter} {et~al.}(2012{\natexlab{a}}){Potter}, {Tout}, \&
  {Eldridge}}]{Potter11}
{Potter} A.~T., {Tout} C.~A., {Eldridge} J.~J., 2012{\natexlab{a}}, \mnras,
  419, 748

\bibitem[{{Potter} {et~al.}(2012{\natexlab{b}}){Potter}, {Tout}, \&
  {Eldridge}}]{Potter12}
---, 2012{\natexlab{b}}, \mnras, In Press

\bibitem[{{Reimers}(1975)}]{Reimers75}
{Reimers} D., 1975, Memoires of the Societe Royale des Sciences de Liege, 8,
  369

\bibitem[{{Schmalz} \& {Stix}(1991)}]{Schmalz91}
{Schmalz} S., {Stix} M., 1991, \aap, 245, 654

\bibitem[{{Spruit}(1999)}]{Spruit99}
{Spruit} H.~C., 1999, \aap, 349, 189

\bibitem[{{Spruit}(2002)}]{Spruit02}
---, 2002, \aap, 381, 923

\bibitem[{{Stancliffe} \& {Eldridge}(2009)}]{Eldridge09}
{Stancliffe} R.~J., {Eldridge} J.~J., 2009, \mnras, 396, 1699

\bibitem[{{Talon} {et~al.}(1997){Talon}, {Zahn}, {Maeder}, \&
  {Meynet}}]{Talon97}
{Talon} S., {Zahn} J.-P., {Maeder} A., {Meynet} G., 1997, \aap, 322, 209

\bibitem[{{Tayler}(1973)}]{Tayler73}
{Tayler} R.~J., 1973, \mnras, 161, 365

\bibitem[{{Tout} \& {Pringle}(1996)}]{Tout96}
{Tout} C.~A., {Pringle} J.~E., 1996, \mnras, 281, 219

\bibitem[{{ud-Doula} \& {Owocki}(2002)}]{ud-Doula02}
{ud-Doula} A., {Owocki} S.~P., 2002, \apj, 576, 413

\bibitem[{{Vink} {et~al.}(2001){Vink}, {de Koter}, \& {Lamers}}]{Vink01}
{Vink} J.~S., {de Koter} A., {Lamers} H.~J.~G.~L.~M., 2001, \aap, 369, 574

\bibitem[{{Wade} {et~al.}(2009){Wade}, {Alecian}, {Bohlender}, {Bouret},
  {Grunhut}, {Henrichs}, {Neiner}, {Petit}, {Louis}, {Auri{\`e}re},
  {Kochukhov}, {Silvester}, {ud-Doula}, \& {ud-Doula}}]{Wade09}
{Wade} G.~A., {Alecian} E., {Bohlender} D.~A., {Bouret} J.-C., {Grunhut} J.~H.,
  {Henrichs} H., {Neiner} C., {Petit} V., {Louis} N.~S., {Auri{\`e}re} M.,
  {Kochukhov} O., {Silvester} J., {ud-Doula} A., {ud-Doula}, 2009, in IAU
  Symposium, Vol. 259, IAU Symposium, pp. 333--338

\bibitem[{{Yousef} {et~al.}(2003){Yousef}, {Brandenburg}, \&
  {R{\"u}diger}}]{Yousef03}
{Yousef} T.~A., {Brandenburg} A., {R{\"u}diger} G., 2003, \aap, 411, 321

\bibitem[{{Zahn}(1992)}]{Zahn92}
{Zahn} J.-P., 1992, \aap, 265, 115

\end{thebibliography}

\end{document}